\documentclass[twocolumn]{aastex63}
\pdfoutput=1 
\usepackage{amsmath,amstext}
\usepackage[T1]{fontenc}
\usepackage{apjfonts}
\usepackage[figure,figure*]{hypcap}
\usepackage{upgreek}
\usepackage{float}
\usepackage{rotating}
\usepackage{listings}
\usepackage{hyperref}
\usepackage{xcolor}

\usepackage{xspace}
\newcommand{\Teff}{$T_\mathrm{eff}$\xspace}
\newcommand{\Lbol}{$L_\mathrm{bol}$\xspace}
\usepackage{booktabs}
\usepackage{threeparttable}
\usepackage{threeparttablex}
\usepackage{subfloat}
\usepackage{lineno}

\graphicspath{{./}{figures/}}

\received{June 13, 2022}
\revised{September 1, 2022}
\accepted{September 1, 2022}
\submitjournal{ApJ}

\shorttitle{P-T Profile Comparisons}
\shortauthors{Gonzales et al.}

\begin{document}

\title{A Comparative L-dwarf Sample Exploring the Interplay Between Atmospheric Assumptions and Data Properties}
 

\correspondingauthor{Eileen Gonzales}
\email{ecg224@cornell.edu}

\author[0000-0003-4636-6676]{Eileen C. Gonzales}
\altaffiliation{51 Pegasi b Fellow}
\affiliation{Department of Astronomy and Carl Sagan Institute, Cornell University, 122 Sciences Drive, Ithaca, NY 14853, USA}
\affiliation{Department of Astrophysics, American Museum of Natural History, New York, NY 10024, USA}

\author[0000-0003-4600-5627]{Ben Burningham}
\affiliation{Centre for Astrophysics Research, School of Physics, Astronomy and Mathematics, University of Hertfordshire, Hatfield AL10 9AB}

\author[0000-0001-6251-0573]{Jacqueline K. Faherty}
\affiliation{Department of Astrophysics, American Museum of Natural History, New York, NY 10024, USA}

\author[0000-0002-8507-1304]{Nikole K. Lewis}
\affiliation{Department of Astronomy and Carl Sagan Institute, Cornell University, 122 Sciences Drive, Ithaca, NY 14853, USA}

\author[0000-0001-6627-6067]{Channon Visscher}
\affiliation{Chemistry \& Planetary Sciences, Dordt University, Sioux Center, IA}
\affiliation{Center for Extrasolar Planetary Systems, Space Science Institute, Boulder, CO}

\author[0000-0002-5251-2943]{Mark Marley}
\affiliation{Department of Planetary Sciences and Lunar and Planetary Laboratory, University of Arizona, Tuscon, AZ}

\begin{abstract}
Comparisons of atmospheric retrievals can reveal powerful insights on the strengths and limitations of our data and modeling tools. In this paper, we examine a sample of 5 similar effective temperature (\Teff) or spectral type L dwarfs to compare their pressure-temperature (P-T) profiles. Additionally, we explore the impact of an object's metallicity and the observations' signal-to-noise (SNR) on the parameters we can retrieve. We present the first atmospheric retrievals: 2MASS J15261405$+$2043414, 2MASS J05395200$-$0059019, 2MASS J15394189$-$0520428, and GD 165B increasing the small but growing number of L-dwarfs retrieved. When compared to atmospheric retrievals of SDSS J141624.08+134826.7, a low-metallicity d/sdL7 primary in a wide L+T binary, we find similar \Teff sources have similar P-T profiles with metallicity differences impacting the relative offset between their P-T profiles in the photosphere. We also find that for near-infrared spectra, when the SNR is $\gtrsim80$ we are in a regime where model uncertainties dominate over data measurement uncertainties. As such, SNR does not play a role in the retrieval's ability to distinguish between a cloud-free and cloudless model, but may impact the confidence of the retrieved parameters. Lastly, we also discuss how to break cloud model degeneracies and the impact of extraneous gases in a retrieval model. 
\end{abstract}


\section{Introduction} \label{sec:intro}
Brown dwarfs form a crucial link between stars and planets, with masses ranging from $\sim13-75\, M_\mathrm{Jup}$ and temperatures of $\sim 250-3000$~K \citep{Saum96, Chab97}. With masses below the hydrogen burning limit, their cores never reach temperatures and pressure required for nuclear fusion and instead are sustained by electron degeneracy pressure \citep{Kumar63a, Haya63}. As such, as brown dwarfs age they continuously cool morphing through the spectral sequence of M, L, T, and Y defined by their optical and/or near-infrared (NIR) spectra \citep{Burg02a, Kirk05,Cush11}. This lack of sustained stable hydrogen burning in the cores of brown dwarfs leads to a degeneracy between their effective temperature (\Teff), mass, and age making these parameters difficult to disentangle. 

One way to break the degeneracy is by obtaining the kinematics of an object along with its spectrum. Young brown dwarfs can display kinematics placing them in nearby young moving groups (e.g.  \citealt{Fahe16, Kell16, Gagn17}) or spectral features indicative of youth such as weak alkali lines and enhanced metal oxides \citep{Kirk06,Cruz09,Alle10}. Subdwarfs display unusually blue NIR colors, metallicities lower than that of the Sun, kinematics placing them in the galactic halo, and spectral features such as weak or absent metal oxides and enhanced metal hydrides \citep{Burg03c,Burg09a,Gizi97, Dahn08,Burg08a, Cush09}. This has enabled the separation of brown dwarfs into roughly three main age subpopulations-- field, young (low-gravity), and subdwarfs (low-metallicity), corresponding to ages of roughly $0.5-10$~Gyr (e.g. \citealt{Fili15}), $<500$~Myr (e.g. \citealt{Bell15}), and $>5$~Gyr (e.g. \citealt{Gonz18}).

With age groupings determined, many works have examined observational correlations between the bolometric luminosity (\Lbol), \Teff, spectral type, and the colors of these subpopulations. When comparing NIR color and spectral type, low-gravity sources appear redder than field sources, while subdwarfs appear bluer. Observational studies have lead to the creation of polynomial relations for \Lbol, \Teff, and magnitude versus spectral type for field, low-gravity sources, and subdwarfs (e.g. \citealt{Fili15, Fahe16, Dupu17, Zhang2017a}), as well as the development of gravity sensitive indices \citep{Alle13}. 

As another way to probe the Mass-\Teff-Age degeneracy, comparisons of spectral energy distributions (SED) and the resultant semi-empirical fundamental parameters have been made across objects of similar \Lbol and \Teff (e.g. \citealt{Fili15, Gonz18, Gonz19}). Comparisons highlight the differences in overall SED shape, as well as the impact clouds, metallicity, and surface gravity play in shaping those differences in NIR spectral bands. For example, \cite{Gonz18} highlights the impact clouds have on the overall SED shape for objects of the same \Teff but differing ages. They show that cloudy young objects redistribute flux out to longer wavelengths while low-metallicity objects do not, due to a lack of clouds in their atmospheres. 

To more deeply compare the atmospheres of brown dwarfs, we need to move to using a theoretical approach. In this work we use atmospheric retrievals, an in-depth data-driven spectral modeling approach to explore the physical and chemical processes occurring in an object's atmosphere. These processes impact an object's chemical composition, cloud properties, formation, and evolution. By using atmospheric retrievals, we can compare the similarities and differences between the thermal structure of brown dwarfs of similar \Teff or spectral type. At present, retrievals of brown dwarfs are still a relatively new approach in the field. There are a growing number of L, T, and Y dwarfs examined (e.g. \citealt{Line14, Line15,Line17, Zale19, Kitz20, Burn17, Burn21, Gonz20, Gonz21,Piet20,Lueb22, Howe22}), however, none of these studies to date have completed a comparison between the retrieved properties with a sample of objects anchored by their observational and fundamental parameter similarities such as what we present in this work. 

Using the \textit{Brewster} retrieval framework \citep{Burn17} we pose the question ``How do the pressure-temperature (P-T) profiles of objects of similar spectral type or similar \Teff compare to one another?'' To address this we created a sample anchored by the d/sdL7 dwarf SDSS J141624.08+134826.7 (hereafter J1416$+$1348A) with one optical spectral type comparison and three \Teff comparisons based on the \Teff from their SEDs. We chose multiple \Teff comparisons for J1416$+$1348A as its \Teff differs by $\sim200$~K between SED and retrieval methods \citep{Gonz20}. In Section~\ref{sec:litdata} we present literature data on the sample. We present semi-empirical fundamental parameters for the sample generated via SEDs in Section~\ref{sec:Fundparams}. We discuss the \textit{Brewster} retrieval code framework in Section~\ref{sec:RetrievalModel} and present our updates to the \cite{Gonz20} retrieval of J1416$+$1348A in Section~\ref{sec:1416Aupdate}. Sections~\ref{sec:modelselection} and \ref{sec:RetrievalResults} describe model section and results for our comparative sample. Section~\ref{sec:discussion} compares the P-T profiles for the spectral type and \Teff samples, discusses the impact of wavelength coverage and metallicity on the shape of the P-T profile, and the impact of signal-to-noise (SNR) on model selection and retrieved property constraints.


\section{Literature Data on Sample}\label{sec:litdata}
Relevant literature data for the sample is presented in the following subsections and listed in Table~\ref{tab:Litdata1416comp}.

\begin{deluxetable*}{l c c c c c c c c cc} 
\tabletypesize{\footnotesize}
\tablecaption{Selected Literature Data on the Comparative Sample \label{tab:Litdata1416comp}} 
\tablehead{\colhead{} & \colhead{} & \colhead{} & \multicolumn{2}{c}{SpT Comp} & \multicolumn{6}{c}{\Teff Comp}\\
           \cmidrule(lr){4-5}\cmidrule(lr){6-11}
           \colhead{Property} & \colhead{J1416$+$1348A} & \colhead{Ref.} & \colhead{J1526$+$2043} & \colhead{Ref.} & \colhead{J0539$-$0059} & \colhead{Ref.} & \colhead{J1539$-$0520} & \colhead{Ref.} & \colhead{GD 165B} & \colhead{Ref.}} 
  \startdata
  Optical Spectral Type & d/sdL7   & 1        & L7 & 2 & L5 & 3 & L4 & 4 & L4 & 5 \\
  NIR Spectral Type     & L6.0pec$\pm2.0$ & 6 & L5 & 7 & L5 & 8 & L2 & 9 & L3 & 8\\ \hline
  \multicolumn{11}{c}{Astrometry} \\ \hline
  R.A.                         & $14^h 16^m 24.08^s$    & 10 & $15^h 26^m 14.05^s$     & 10 & $05^h 39^m 52.00^s$    & 10 & $15^h 39^m 41.89^s$    & 10 & $14^h 24^m 39.09^s$   & 10\\
  Decl.                        & $+13 ^\circ 48' 26''.3$& 10 & $+20 ^\circ 43' 41.4''$ & 10 & $-00^\circ 59' 01.9''$ & 10 & $-05^\circ 20' 42.8''$ & 10 & $+09^\circ 17' 10.4''$& 10 \\
  $\pi$ (mas)                  & $107.56 \pm 0.30$      & 11 & $50.00 \pm 1.48$       & 11 & $78.53\pm0.57$        & 11 & $58.82\pm0.42$        & 11 & $29.933\pm0.06$  &11\\
  $\mu_\alpha$ (mas yr$^{-1}$) & $85.69\pm0.69$         & 11 & $-220.78\pm 2.34$      & 11 & $162.45\pm0.79$       & 11 & $590.20\pm0.75$       & 11 & $-213.35\pm0.08$ &11\\              
  $\mu_\delta$ (mas yr$^{-1}$) & $129.07\pm0.47$        & 11 & $-359.16\pm 2.04$      & 11 & $321.15\pm0.76$       & 11 & $104.57\pm0.73$       & 11 & $-149.65\pm0.07$ &11\\
  $vsin\,i$                    & $\cdots$               & $\cdots$                    & $\cdots$            & $\cdots$. & $32.30 \pm 0.75$      & 12 & $40.09\pm 0.76$  & 12 & $18\pm2$ & 13\\
  $V_{r}$ (km s$^{-1}$)        & $-42.2 \pm 1.24$       & 14                          & $\cdots$            & $\cdots$  & $13.91 \pm 0.15$      & 12 & $27.33\pm 0.24$  & 12 & $-29\pm2$ & 15\\ \hline 
  \multicolumn{11}{c}{Parameters from SED} \\ \hline
 $L_\mathrm{bol}$           & $-4.18\pm0.011$& 16 & $-4.339 \pm 0.022$ & 17 & $-4.18 \pm 0.016$& 17 & $-3.959 \pm 0.016$ & 17 &$-3.99\pm0.04$ &17\\ 	
  $T_\mathrm{eff}$ (K)      & $1694\pm74$    & 16 & $1491 \pm 78$      & 17 & $1633 \pm 68$    & 17 & $1837 \pm 66$      & 17 &$1804\pm75$    &17\\
  Radius ($R_\mathrm{Jup}$) & $0.92\pm0.08$  & 16 & $0.99 \pm 0.10$    & 17 & $0.99 \pm 0.08$  & 17 & $1.01 \pm 0.07$    & 17 &$1.01\pm0.07$  &17\\ 
  Mass ($M_\mathrm{Jup}$)   & $60\pm18$      & 16 & $56 \pm 18$        & 17 & $60 \pm 15$      & 17 & $65 \pm 13$        & 17 &$64\pm14$      &17\\ 
  log $g$ (dex)             & $5.22\pm0.22$  & 16 & $5.12 \pm 0.24$    & 17 & $5.16 \pm 0.2$   & 17 & $5.18 \pm 0.16$    & 17 &$5.18\pm0.17$  &17\\     
  Age (Gyr)                 & $0.5-10$       & 16 & $0.5-10$           & 17 & $0.5-10$         & 17 & $0.5-10$           & 17 &$0.5-10$       &17 \\
  Distance \textbf{(pc)}    & $9.3\pm0.3$    & 16 & $20.0\pm 0.6$      & 17 & $12.73\pm 0.09$  & 17 & $17.0\pm0.1$       & 17 &$33.41\pm0.06$ &17 \\ \hline
  \multicolumn{11}{c}{Retrieved Parameters\tablenotemark{a}\tablenotemark{b}} \\ \hline
  $L_\mathrm{bol}$          & $-4.22 \pm 0.01$                     & 17 & $-4.35\pm 0.01$                    & 17 & $-4.21\pm{0.01}$                  & 17 & $-3.94\pm 0.01$                   & 17 & $-4.05\pm 0.01$                   & 17 \\
  $T_\mathrm{eff}$ (K)      & $1892.31\substack{+34.48 \\ -39.26}$ & 17 & $1708.33\substack{+20.12\\-19.82}$ & 17 & $1859.57\substack{+36.00\\-33.52}$& 17 & $1926.13\substack{+29.87\\-32.44}$& 17 & $1920.49\substack{+27.88\\-27.93}$& 17 \\
  Radius ($R_\mathrm{Jup}$) & $0.7 \substack{+0.4 \\ -0.3}$        & 17 & $0.74\pm{0.06}$                    & 17 & $0.74\pm{0.04}$                   & 17 & $0.93\pm{0.04}$                   & 17 & $0.83\pm{0.04}$                   & 17 \\ 
  Mass ($M_\mathrm{Jup}$)   & $36.36\substack{+28.4 \\ -19.0}$     & 17 & $76.57\substack{+20.21\\-25.24}$   & 17 & $34.15\substack{+30.74\\-18.19}$  & 17 & $53.98\substack{+31.69\\-24.79}$  & 17 & $43.96\substack{+34.62\\-23.24}$  & 17 \\
  log $g$ (dex)             & $5.25\substack{+0.27 \\ -0.35}$      & 17 & $5.55\substack{+0.07\\-0.18}$      & 17 & $5.20\substack{+0.29\\-0.36}$     & 17 & $5.19\substack{+0.21\\-0.28}$     & 17 & $5.20\substack{+0.26\\-0.34}$     & 17 \\     
  C/O\tablenotemark{c}      & $0.62\substack{+0.08 \\ -0.10}$      & 17 & $0.74\pm0.04$                      & 17 & $0.71\substack{+0.07 \\ -0.05}$   & 17 & $0.63\substack{+0.08 \\ -0.10}$   & 17 & $0.80\pm{0.05}$                   & 17 \\
  $[M/H]$\tablenotemark{d}  & $-0.16\substack{+0.17 \\ -0.19}$     & 17 & $0.34\substack{+0.12 \\ -0.11}$    & 17 & $0.11\substack{+0.16 \\ -0.15}$   & 17 & $-0.08\substack{+0.14 \\ -0.17}$  & 17 & $0.44\pm{0.18}$                   & 17 \\
  \enddata
  \tablenotetext{a}{\Lbol, \Teff, radius, mass, and {[M/H]} are not directly retrieved parameters, but are calculated using the retrieved $R^2/D^2$ and log\,$g$ values along with the predicted spectrum.}
  \tablenotetext{b}{Values are shown for the ``winning" models. SDSS J1416$+$1348A: Power-law deck cloud, uniform-with-altitude gas abundances. J1526$+$2043: Cloud-free, uniform-with-altitude gas abundances. For both J0536$-$0059 and J1539$-$0520 the power-law deck cloud, uniform-with-altitude gas abundances and cloud free, uniform-with-altitude gas abundances models were indistinguishable. We list values for the cloudy model as it has the higher maximum likelihood. See Table \ref{tab:RetrievalParams} for cloudless model values. GD 165B: Power-law deck cloud, uniform-with-altitude gas abundances.} 
  \tablenotetext{c}{Absolute inferred atmospheric C/O calculated using H$_2$O, CO, and VO abundances. Solar C/O=0.55.}
  \tablenotetext{d}{Inferred metallicity determined using H$_2$O, CO, VO, CrH, FeH, and Na+K abundances. Values are relative to Solar.}
  \tablecomments{In our retrieval models distances and corresponding uncertainties are scaled to 10 pc. Therefore we set the distance to the following J1416$+$1348A:$10\pm0.58$, J1526$+$2043:$10\pm0.6$, J0539$-$0059:$10\pm0.09$, J1539$-$0520:$10\pm0.12$, GD 165B:$10\pm0.02$.}
\tablerefs{(1) \cite{Burn10}, (2) \cite{Kirk00}, (3) \cite{Fan_00}, (4) \cite{Kirk08}, (5) \cite{Kirk99-GD165B}, (6) \cite{Bowl10}, (7)\cite{Schn14}, (8) \cite{Geba02}, (9) \cite{Kend04}, (10) \cite{Cutr03}, (11) \cite{GaiaDR1,GaiaDR2,Lind18}, (12) \cite{Blak10}, (13) \cite{Moha03}, (14) \cite{Schm10a}, (15) \cite{Prat15}, (16) \cite{Gonz20}, (17) This Paper.}
\tabletypesize{\normalsize}
\end{deluxetable*}

\subsection{J1416$+$1348A}
J1416$+$1348A was independently discovered by \cite{Burn10,Schm10a} and \cite{Bowl10}. It is spectral typed as d/sdL7 in the optical \citep{Burn10} due to its unusually blue NIR color and spectral features implying a low-metallicity and/or high surface gravity \citep{Schm10a}. There are currently three optical spectra (\citealt{Schm10a} SDSS and MagE and \citealt{Kirk16} Palomar), 3 NIR spectra (Spex Prism: \citealt{Schm10a, Bowl10}, Spex SXD: \citealt{Schm10a}), and one L band spectrum \citep{Cush10} available of J1416$+$1348A. $Gaia$ DR2 provides parallax and proper motion measurements \citep{GaiaDR1,GaiaDR2,Lind18} and SDSS DR7 provides radial velocity measurements \citep{Abaz09,Schm10a,Bowl10}. Fundamental parameters for J1416$+$1348A have been determined by fitting its self-consistent grid models to its spectrum \citep{Burn10,Schm10a, Bowl10, Cush10} as well as through generating its SED and atmospheric retrievals using the \textit{Brewster} code base \citep{Gonz20}. Table~\ref{tab:Litdata1416comp} lists all the pertinent parameters.

\subsection{J1526$+$2043, Optical Spectral Type Comparison}
2MASS J15261405$+$2043414 (hereafter J1526$+$2043) was discovered by \cite{Kirk00} and spectral typed as L7 in the optical \citep{Kirk00} and L5 in the NIR \citep{Schn14}. The proper motion of J1526$+$2043 has been measured by \cite{Shep08}, \cite{Fahe09}, and \textit{Gaia} DR2 \citep{GaiaDR1,GaiaDR2,Lind18}. The parallax has been measured by \cite{Fahe12} and \textit{Gaia} DR2. Spectra for J1526$+$2043 cover wavelengths from $\sim0.6$ to $12\,\mu$m (optical: \citealt{Kirk00}, NIR: \citealt{Burg04b}, and mid-infrared (MIR): \citealt{Cush06b}. \Lbol, \Teff, mass, radius, and log$g$ for J1526$+$2043 were initially derived in \cite{Fili15} via SED generation and are updated in this work. See Table~\ref{tab:Litdata1416comp} for all relevant parameters of J1526$+$2043 for this study.

\subsection{J0539$-$0059, \Teff Comparison}
2MASS J05395200$-$0059019 (hereafter J0539$-$0059) was discovered by \cite{Kirk00} and spectral typed as L5 in the optical \citep{Fan_00} and L5 in the NIR \citep{Geba02}. Spectra (optical: \citealt{Fan_00}, NIR: \citealt{Cush05,Schn14}, and MIR: \citealt{Cush06b, Yama10}) and photometry for J0539$-$0059 cover $\sim0.45$ to $12\,\mu$m. It's parallax was measured by \cite{Vrba04} and \textit{Gaia} DR2 \citep{GaiaDR1,GaiaDR2,Lind18}. Radial velocity and $v$sin$\,i$ were measured in \cite{Blak10}, which are listed in Table~\ref{tab:Litdata1416comp}. Fundamental parameters (\Lbol, \Teff, mass, radius, and log$g$) were determined by \cite{Fili15} and are updated in this work using the same method. 

\cite{Bail01} found strong evidence of variability in the $I$ band for J0539$-$0059, with a period of $13.3\pm1.2$ hours, which was noted as a possible the rotation period. Some evidence for spectral variability due to dust was noted in \cite{Bail08}. However, J0539$-$0059 was found to have marginal evidence for long time-scale variability, but no signs of rapid brightness changes in \cite{Koen13}. Furthermore, \cite{Buen14} found no signs of variability in the $J$ and $H$ bands. Values for J0539$-$0059 from the literature and this work are listed in Table~\ref{tab:Litdata1416comp}.

\subsection{J1539$-$0520, \Teff Comparison}
2MASS J15394189$-$0520428 (hereafter J1539$-$0520) was discovered by \cite{Kirk00} and spectral typed as L4 in the optical \citep{Kirk08} and L2 in the NIR \citep{Kend04}. \cite{Burg15} noted its unusually blue NIR color and determined it to be a member of the thin/thick disk population. Optical and NIR spectra are available for J1539$-$0520 \citep{Reid08b, Bard14}. Available photometry covers the optical, NIR, and MIR wavelengths, see Table \ref{tab:photometrySED}. Radial velocity and $vsini$ were measured in \cite{Blak10} and are listed in Table~\ref{tab:Litdata1416comp}. Fundamental parameters for J1539$-$0520 were initially derived in \cite{Fili15} and are updated here using the same method.

J1539$-$0520 was tested for $I$ band linear polarization in \cite{Zapa05}, but no significant evidence of polarization was found. In \cite{Koen13}, $I$ band photometry of J1539$-$0520 was found to have large changes in its nightly mean magnitude, which could be evidence of longer scale variability. However, in the $R$ band, they found the magnitude to be consistent or at least less variable. They also noted no evidence for short time-scale variability. \cite{Miles17} reported J1539$-$0520 as an $I$ band variable source with periodic patterns easily seen by eye in the light curve. The variability was posed to be most likely due to rotation causing an inhomogeneous atmosphere. \cite{Miles17} determined a rotation period of $2.51 \substack{+1.60 \\ -0.55}$ hours and a rotational velocity between $26-65$ km s$^{-1}$. With the rotational velocity and observed $v$sin\,$i$ being nearly equal \cite{Miles17} argued it implied J1539$-$0520 is seen face on. Parameter values for J1539$-$0520 are listed in Table~\ref{tab:Litdata1416comp}.

\subsection{GD 165B, \Teff Comparison} 
GD 165B (also known as 2MASS J14243909+0917104), was the first L dwarf discovered \citep{Beck88} and is a wide spatially resolved companion to the 1.2-5.5 Gyr white dwarf GD 165A \citep{Kirk99-GD165B}. GD 165B has an optical spectral type of L4 and a NIR spectral type of L3 \citep{Kirk99-GD165B,Geba02}. It has numerous optical spectra (RC Spec:\citealt{Kirk93}, LRIS:\citealt{Kirk99-GD165B}, HIRES: \citealt{Moha03}) and NIR spectra (SpeX prism:\citealt{Bard14}, NIRSPEC:\citealt{McLe07,Prat15, Martin17}) as well as photometric measurements listed in Table~\ref{tab:photometrySED}. Astrometric measurements of GD 165B include the parallax \citep{Tinn95,vanA95,GaiaDR1,GaiaDR2,Lind18}, tangential velocity \citep{Tinn95}, radial velocity \citep{Prat15}, and $v$sin\,$i$ \citep{Moha03}. The most precise values of these are listed in Table~\ref{tab:Litdata1416comp}. Fundamental parameters for GD 165B are derived in \cite{Legg02a} and \cite{Fili15}. We update them in this work using \cite{Fili15} method.

\section{Fundamental Parameters and overall SED Comparison of the Sample}\label{sec:Fundparams}
We used the SED for J1416$+$1348A from \cite{Gonz20} and for each comparative object generated their SEDs using the \cite{Fili15} method (and code SEDkit)\footnote{SEDkit is available on GitHub at \url{https://github.com/hover2pi/SEDkit}. The Eileen branch was used for this work (also available at \url{https://github.com/ECGonzales/SEDkit/tree/master}).}, which we briefly discuss below for our targets. For each comparative source except GD 165B, as we only have its NIR data, the optical and NIR spectra (and MIR when available) were combined into a composite spectrum, with the average spectra taken in regions of overlap. In regions with no overlap, linear interpolation was used to connect the spectra. The composite spectrum was then scaled to the absolute magnitudes of the observed photometry (see \cite{Fili15} Section 4.2 for details). \Lbol was determined by integrating under the distance-calibrated SED from 0 to 1000~$\mu$m. As done in \cite{Fili15}, to obtain a radius estimate and the \Teff, we used \cite{Saum08} hybrid cloud evolutionary models  as well as the \cite{Chab00} and \cite{Saum08} $f_{sed}=2$ evolutionary models to determine the radius ranges for on all comparative sources. Additionally due to the later spectral type of J1526$+$2043, the \cite{Bara03} and cloud-free \cite{Saum08} evolutionary models were also used as done in \cite{Fili15}. The final radius range was then set as the maximum and minimum from all model predictions. For all models we used an age range of $0.5-10$ Gyr, corresponding to the field ultracool dwarf age range \citep{Fili15}. The \Teff was calculated using the Stefan--Boltzmann law along with the inferred radius and calculated \Lbol. The range of masses was determined using the same evolutionary models used to determine the radius. The photometry and spectra used in the creation of the SEDs are listed in Tables~\ref{tab:photometrySED}--\ref{tab:SpectraSED} in Appendix~\ref{sec:Appendix} and the fundamental parameters we derive are in Table \ref{tab:Litdata1416comp}.

\begin{figure*} 
 \centering
  \includegraphics[scale=0.65]{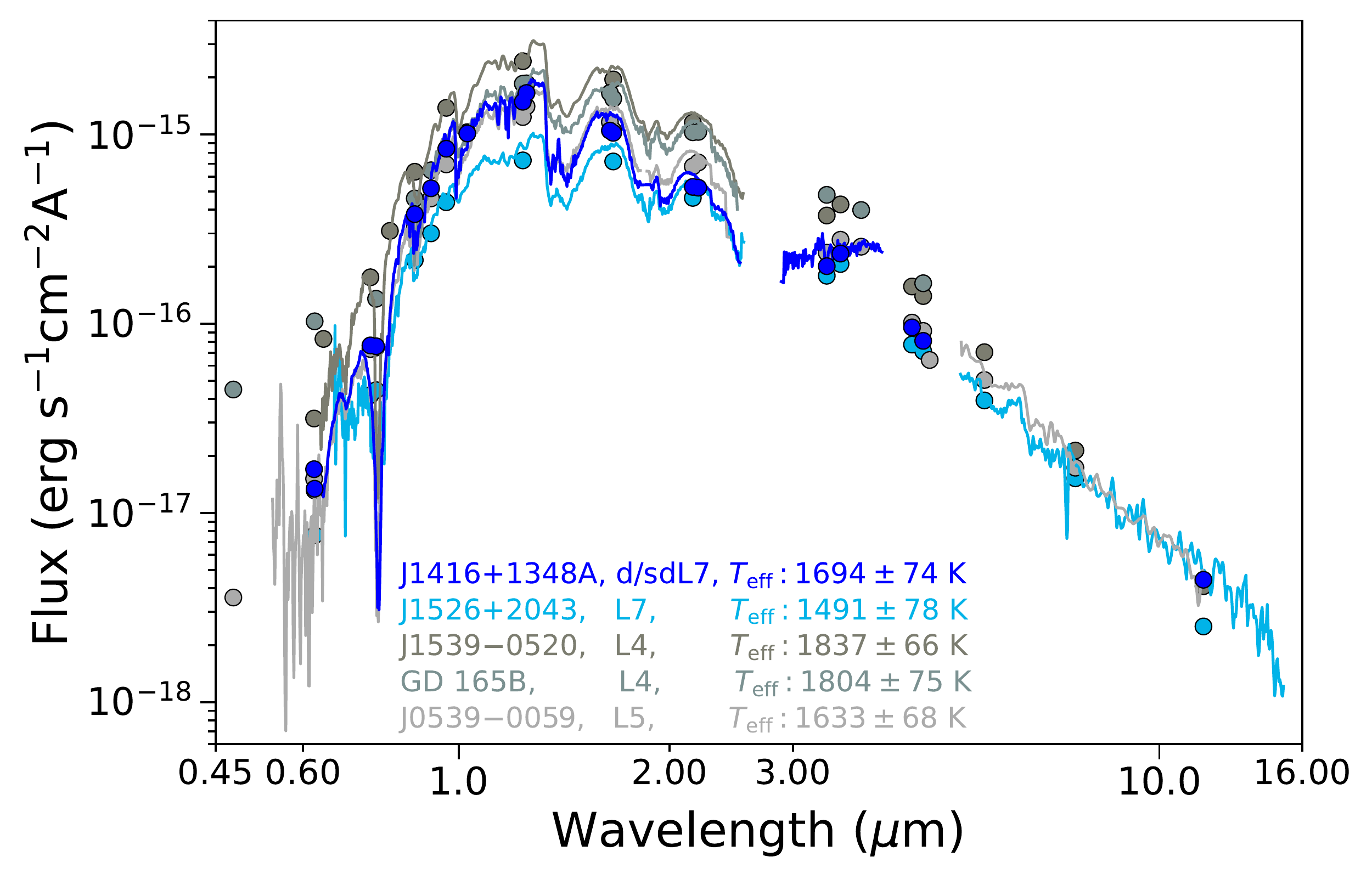}
\caption[Distance-calibrated SEDs of J1416$+$1348A, J1526$+$2043, J1539$-$0520, GD 165B, and J0539$-$0059.]{Distance-calibrated SEDs of J1416$+$1348A (dark blue) and the comparative sources of similar \Teff (J0539$-$0059, GD 165B, and J1539$-$0520 in greys) and of the same optical spectral type (J1526$+$2043 in light blue). All spectra were resampled to the same dispersion relation using a wavelength-dependent Gaussian convolution. The spectra are flux calibrated and scaled to the absolute magnitudes of the photometry shown. No normalization is applied.}
\label{fig:1416compSED}
\end{figure*}

The SEDs of the comparative sample along with J1416$+$1348A across $0.45-16\, \mu$m are shown in Figure~\ref{fig:1416compSED}. The SED of J1539$-$0520 is overluminous across all wavelengths when compared to J1416$+$1348A. For the \Teff comparisons except J1539$-$0520, we see that they overlap with one another and J1416$+$1348A in the optical, while they have a large spread in flux ``fanning out" in the NIR. This is derives from the differences in L dwarf spectra, where the optical spectral type classification is anchored by the \Teff with the NIR showing diversity from secondary parameters \citep{Cruz18}. As discussed in \cite{Zhang2017a} and \cite{Gonz18}, subdwarfs have effective temperatures similar to objects 2 to 3 spectral types earlier and thus J1416$+$1348A agrees well in the optical to the \Teff comparisons due to this. We find that J0539$-$0059, the coolest \Teff comparison, most closely matches the overall shape of J1416$+$1348A from the optical through mid $H$ band. GD 165B is a close match to J1416+1348A from $\sim0.85-1.33\, \mu$m, but beyond it is over luminous. In the $K$ band and beyond, J1416A more closely shares similarities to the optical spectral type comparison J1526$+$2043. 

It should be noted that as J1539$-$0520 is a known variable source, this SED does not represent a consistent picture of the atmosphere at a single point in time, as the data have been taken on different dates. This could also be true for J0539$-$0059 as it may be variable. As a test to see if variations on the order of 20\% make a difference in the fundamental parameters, adding/subtracting the photometric uncertainty in each band to the measured photometry would produce a variation near this level. From this we find that all fundamental parameters remain within the uncertainties, thus variability at this level should not affect the derived fundamental parameters.

\section{The \textit{Brewster} Retrieval Framework \label{sec:RetrievalModel}} 
Our retrievals use the \textit{Brewster} framework \citep{Burn17} with a modified setup similar to \cite{Gonz20}. The \textit{Brewster} framework is described in extensive detail in \cite{Burn17}, \cite{Gonz20}, and \cite{Gonz21}. Below we briefly describe key components in the \textit{Brewster} framework and our modifications for this work.

As done in \cite{Gonz20}, we use the same forward model with the \cite{Madh09} five parameter pressure-temperature (P-T) profile, gas opacities (only using the Allard opacites for the alkalies), both the uniform-with-altitude and thermochemical equilibrium methods for determining gas abundances, and the same cloud models. The cloud model has the option of either a ``slab" or ``deck" cloud. Both clouds are defined with the opacity distributed over pressure layers with the optical depth as either grey or a power-law ($\tau = \tau_0\lambda^\alpha$, where $\tau_0$ is the optical depth at 1~$\mu$m). The deck cloud defined to be optically thick at some pressure level such that the vertical extent of the cloud at lower pressures can only be seen. Three parameters define the deck cloud: (1) the cloud top pressure $P_{top}$, the point at which the cloud passes $\tau=1$ (looking down), (2) the decay height, $\Delta \log P$, over which the optical depth falls to lower pressures as $d\tau/dP \propto \exp((P-P_{deck}) / \Phi)$ where $\Phi = (P_{top}(10^{\Delta \log P} - 1))/(10^{\Delta \log P})$, and (3) the cloud particle single-scattering albedo, which as done in \cite{Gonz20} is set to zero thereby assuming an absorbing cloud. Unlike the deck cloud, the slab cloud is defined in pressure space to have a top and bottom. Therefore, an additional parameter for is included determining the total optical depth at 1$\mu$m ($\tau_{cloud}$), which is distributed through the slab cloud extent as $d\tau / dP \propto P$ (looking down), reaching its total value at the bottom (highest pressure) of the slab. If the deck or slab cloud has an non-grey opacity, an additional parameter for the power ($\alpha$) in the optical depth is included. For additional details on the cloud model see \citealt{Gonz20, Gonz21}, Section 4.4. The same thermochemical grids are used as in \cite{Gonz20}, which were calculated using the NASA Gibbs minimization CEA code \citep{McBr94}, based on previous thermochemical models \citep{Fegl94,Fegl96,Lodd99,Lodd02,Lodd02b,Lodd10,Lodd06,Viss06,Viss10a,Viss12,Moses12,Moses13} and recently utilized to explore gas and condensate chemistry over a range of conditions in substellar atmospheres \citep{Morl12,Morl13,Skem16,Kata16,Wake17,Gonz20,Burn21,Ghar21, Gonz21, Sonora, Kara21}. Chemical grids in this work determine equilibrium abundances of atmospheric species over temperatures between $300-4000$~K, pressures between 1 microbar to 300 bar, metallicities ranging from $-1.0 < [\mathrm{Fe/H}] < +2.0$, and C/O abundance ratios of $0.25$ to $2.5$x the solar abundance. 

\subsection{Retrieval Model}\label{sec:Retmodelsetup} 
Our retrieval model uses the EMCEE package \citep{emcee} to sample posterior probabilities with the priors we used shown in Table \ref{tab:Priors}. As done in \cite{Gonz20}, for all objects in the sample we extend the mass and temperature priors to allow for surface gravities that encompass the possible ranges for subdwarfs. Distance-calibrated SpeX prism spectra (output from generating our SED) are trimmed to $1.0-2.5\,\mu$m to remove the $0.8-1.0\,\mu$m region of the spectrum impacted by the pressure bordering from the 0.77$\mu$m \ion{K}{1} doublet and set the distance to 10~pc with the correspondingly scaled uncertainty for our retrieval as done in \cite{Gonz20}.

\begin{deluxetable*}{l c c}
\tablecaption{Priors for retrieval models\label{tab:Priors}} 
\tablehead{\colhead{Parameter} &\phm{stringzzzzzzzz} &\colhead{Prior}}
  \startdata
  gas volume mixing ratio && uniform, log $f_{gas} \geq -12.0$, $\sum_{gas} f_{gas} \leq 1.0$ \\
  thermal profile ($\alpha_{1}, \alpha{2}, P1, P3, T3$) && uniform, $0.0\, \mathrm{K} < T < 6000.0\, \mathrm{K}$\\ 
  scale factor ($R^2/D^2$) && uniform, $0.5\,R_\mathrm{Jup} \leq\, R\, \leq 2.0\,R_\mathrm{Jup}$ \\
  gravity (log\,$g$)&& uniform, $1\,M_\mathrm{Jup} \leq\; gR^2/G\; \leq 100\,M_\mathrm{Jup}$ \\
  cloud top\tablenotemark{a} && uniform, $-4 \leq \mathrm{log}\, P_{CT} \leq 2.3$\\ 
  cloud decay scale\tablenotemark{b} && uniform,$0< \mathrm{log}\,\Delta\, P_{decay}<7$\\
  cloud thickness\tablenotemark{c} && uniform, log\,$P_{CT} \leq\,$log $(P_{CT}+\Delta P)\, \leq2.3$\\
  cloud total optical depth at $1\mu$m && uniform,  $0.0 \geq \tau_{cloud} \geq 100.0$ \\
  wavelength shift && uniform, $-0.01 < \Delta \lambda <0.01 \mu$m \\
  tolerance factor && uniform, log($0.01 \times min(\sigma_{i}^2)) \leq b \leq$ log$(100 \times max(\sigma_{i}^2)) $\\
  \enddata
  \tablenotetext{a}{For the deck cloud this is the pressure where $\tau_{cloud} = 1$, for a slab cloud this is the top of the slab.}
  \tablenotetext{b}{Decay height for deck cloud above the $\tau_{cloud} = 1.0$ level.}
  \tablenotetext{c}{Thickness and $\tau_{cloud}$ only retrieved for slab cloud.}
\end{deluxetable*}

In each model for the gases, surface gravity, $\Delta\lambda$ (the wavelength shift between the model and data), the scale factor ($R^2/D^2$), cloud top pressure and power-law parameters we initialize 16 walkers per parameter in a tight Gaussian. Gases are centered around the approximate solar composition equilibrium chemistry values for gas volume mixing ratios. Surface gravity is initialized centered around the SED-derived value. For our tolerance parameter, optical depth, and cloud thickness we use a flat distribution. Our thermal profile is initialized using the \cite{Saum08} \Teff$=1700$~K log\,$g=5.0$ model. To be sure of convergence, each model is run for at least 50 times the autocorrelation length with the EMCEE chain having between 90,000-120,000 iterations. Additionally, we examine the maximum likelihood to ensure that it is no longer fluctuating.

Initially as done in \cite{Gonz20}, we retrieved abundances for the following gases: H$_2$O, CO, CO$_2$, CH$_4$, TiO, VO, CrH, FeH, K, and Na, with K and Na tied together as a single element in the state-vector assuming a Solar ratio taken from \cite{Aspl09}. However, due to the wavelength coverage and temperatures of of sources we were unable to constrain CO$_2$, CH$_4$, and TiO for all objects in our sample. For these sources CO$_2$ and CH$_4$ are not visible in the NIR spectrum, but do play a role at longer wavelengths. However, the abundances we retrieve when including CO$_2$, CH$_4$, and TiO the abundances push up gainst our lower prior bounds and are on the order of $10^{-7.8}$ (for CO$_2$ and CH$_4$) and less than $10^{-9.8}$ (for TiO), much lower than are primary gases H$_2$O and CO. Without these species we can still get meaningful results with only a minor change in our abundances of FeH and CrH. Therefore we removed the unconstrained gases and ran models only including H$_2$O, CO, VO, CrH, FeH, K, and Na which are reported here. Our retrieval model also includes H$^-$ bound-free and free-free continuum opacities to account for the possibility of the profile going above 3000K in the photosphere and are set based on temperature and pulled from the thermochemical equilibrium grids. We test various cloud parameterizations by building up from cloudless to the 4 parameter power-law slab cloud model and test the uniform-with-altitude and thermochemical equilibrium gas abundance methods for all cloud models.

\section{Updating the J1416+1348A Retrieval}\label{sec:1416Aupdate}
By removing the unconstrained gases from our retrieval models for J1416+1348A, we were able to break the degeneracy between the two indistinguishable models in \cite{Gonz20}-- the power-law deck cloud model and the power-law slab cloud model. With CO$_2$, CH$_4$, and TiO removed, the power-law deck cloud model became strongly preferred over the slab cloud model for J1416+1348A. We also find the median P-T profile for the deck cloud model has shifted to become warmer at slightly lower pressures in the photosphere and deeper compared to the \cite{Gonz20} P-T profile, now agreeing more closely with the Sonora \citep{Sonora} model grids (see Figure~\ref{fig:PT_profile_comparisons}(c)). All retrieved gas abundances and derived fundamental parameters for the updated deck cloud model are consistent with those in \cite{Gonz20},as the 1$\sigma$ P-T profiles are nearly identical to those in \cite{Gonz20}. Figures for this updated model can be found on Zenodo\footnote{https://doi.org/10.5281/zenodo.7045012}.

\section{Model Selection}\label{sec:modelselection}
Our retrieval models aimed to test which cloud parameterization and gas abundance method is preferred for each of the comparative objects. We explore the nature of clouds in our objects by building up from the least complex cloudless model to the most complex slab cloud model. For every cloud model, we tested the impact of the uniform-with-altitude and the thermochemical equilibrium gas abundance methods. To rank our models we used the Bayesian Information Criterion (BIC), where the lowest BIC is preferred. We use the following intervals from \cite{Kass95} for selecting between two models, with evidence against the higher BIC as:

\begin{itemize}
\setlength\itemsep{-0.5em}
  \item[] $0  < \Delta$BIC $< 2$: no preference worth mentioning;
  \item[] $2  < \Delta$BIC $< 6$: positive;
  \item[] $6  < \Delta$BIC $< 10$: strong;
 \item[] $10  < \Delta$BIC: very strong.
\end{itemize}

\section{Retrieval Results}\label{sec:RetrievalResults}
Table \ref{tab:RetModelscomps} shows all models tested for our comparative sample with their corresponding $\Delta$BIC values. J1416$+$1348A is best fit by a power-law deck cloud when removing the unconstrained gases from the \cite{Gonz20} retrieval. The spectral type comparison, J1526$+$2043, is best fit by a cloud-free model. Two of the temperature comparisons, J0539$-$0059 and J1539$-$0520, however have two indistinguishable best fit models, a cloud-free model and power-law deck cloud model. While the temperature comparison GD 165B is best fit by a power-law deck cloud model. Implications of this cloud model degeneracy will be discussed in detail in Section~\ref{sec:PTcompTeff}. Additional figures for each comparative source's ``winning'' model(s) can be found on Zenodo\footnote{https://doi.org/10.5281/zenodo.7045012}.

\begin{deluxetable*}{l c c c c c c c}
\tabletypesize{\small} 
\tablecaption{Retrieval Models of the Comparison Sample \label{tab:RetModelscomps}}
\tablehead{\multicolumn{2}{c}{Model} & \colhead{N Params} &\multicolumn{5}{c}{$\Delta$BIC}\\
\cmidrule(lr){1-2}\cmidrule(lr){4-8}
\colhead{Cloud Case} &\colhead{Gas Method\tablenotemark{a}} &\colhead{} & \colhead{J1416+1348A} &\colhead{J1526$+$2043} & \colhead{J0539$-$0059}& \colhead{J1539$-$0520} & \colhead{GD 165B}}
  \startdata
  Cloud Free & uniform & 15     & $\cdots$  &\textbf{0}& \textbf{0} & \textbf{0}  & 3\\ 
  Grey Deck  & uniform & 18     & $\cdots$  & 19       & 9          & 11.9        & 18 \\
  Power-law Deck & uniform & 19 & \textbf{0}& 17       & \textbf{2} & \textbf{0.4}& \textbf{0} \\  
  Grey Slab & uniform & 19      & $\cdots$  & 21       & 12         & 18.7        & 23\\
  Power-law Slab & uniform & 20 & 5         & 14       & 12         & 7.7         & 8  \\ \hline
  Cloud Free & CE & 11          & $\cdots$  & 12       & 36         & 25.7        & 8  \\ 
  Grey Deck & CE & 14           & $\cdots$  & 4        & 40         & 39.0        & 9  \\
  Power-law Deck & CE& 15       & $\cdots$  & 6        & 5          & 27.1        & 17  \\ 
  Grey Slab & CE & 15           & $\cdots$  & 10       & 12         & 31.1        & 15\\
  Power-law Slab & CE & 16      & $\cdots$  & 11       & 13         & 24.1        & 22\\
  \enddata
  \tablecomments{Winning model BIC in bold for ease.}
  \tablenotetext{a}{Method used to determine gas abundances. uniform = uniform-with-altitude mixing, CE = Chemical Equilibrium.}
\end{deluxetable*}

Retrieved and derived parameters for the cloud-free and power-law deck cloud models of each object can be found in Table~\ref{tab:RetrievalParams}. The derived radius is determined from the parallax value and the retrieved scale factor ($R^2/D^2$). We then calculate the mass using the derived radius and the retrieved log\,$g$. The \Teff is determined using radius and integrating the flux in the resultant forward model spectrum across $0.7-20$~$\mu$m. Lastly, we derive the atmospheric metallicity using the following equations:
\begin{equation}
    f_{H_2}=0.84(1-f_{gases})
\end{equation}
\begin{equation}
    N_{H}=2f_{H_2}N_{tot}
\end{equation}
\begin{equation}
    N_{element}=\sum_{molecules}  n_{atom}f_{molecule}N_{tot}
\end{equation}
\begin{equation}
    N_{M}=\sum_{elements} \frac{N_{element}}{N_H}
\end{equation}
where $f_{H_2}$ is the H$_2$ fraction, $f_{gases}$ is the total gas fraction for all other gases, $N_{H}$ is the number of neutral hydrogen atoms, $N_{tot}$ is the total number of gas molecules, $N_{element}$ is the number of atoms for the element of interest, and $n_{atom}$ is the number of atoms of that element in a molecule (e.g. 2 for oxygen in CO$_2$). Thus the final value of [M/H] is
\begin{equation}
    [M/H]=log\frac{N_M}{N_{Solar}}
\end{equation}
where $N_{Solar}$ is calculated as the sum of the solar abundances (from \citealt{Aspl09}) relative to H. We do not account for gases that are invisible (e.g. Nitrogen) in our atmospheric metallicity calculation. For all comparative sources we find that retrieved and derived parameters are consistent between the cloud-free and cloudy models, with the exception of the \Teff for J1539$-$0520 and GD 165B. Importantly as the C/O ratio and metallicity between both cloudless and cloudy models are consistent, if  gas abundances and fundamental parameters are only of concern, then one can confidently derive these while using a poorer fitting cloud model for theses sources and likely others if they have behave likewise. 

\begin{deluxetable*}{l c c c c c c c c cc}
\tabletypesize{\scriptsize}
\tablecaption{Retrieved Gas Abundances and Derived Properties for the Comparison Sample \label{tab:RetrievalParams}} 
\tablehead{\colhead{} & \colhead{} & \colhead{} & \multicolumn{2}{c}{SpT Comp} & \multicolumn{6}{c}{\Teff Comp}\\
           \cmidrule(lr){6-11}
           \colhead{} & \multicolumn{2}{c}{J1416$+$1348A (d/sdL7)}& \multicolumn{2}{c}{J1526$+$2043 (L7)} & \multicolumn{2}{c}{J0539$-$0059 (L5)}  &\multicolumn{2}{c}{J1539$-$0520 (L4)} &\multicolumn{2}{c}{GD 165B (L4)}\\
           \cmidrule(lr){2-3}\cmidrule(lr){4-5}\cmidrule(lr){6-7}\cmidrule(lr){8-9}\cmidrule(lr){10-11}
           \colhead{Property} & \colhead{Cloud} & \colhead{Cloud-free} & \colhead{Cloud} & \colhead{Cloud-free} & \colhead{Cloud} & \colhead{Cloud-free} & \colhead{Cloud} & \colhead{Cloud-free}& \colhead{Cloud} & \colhead{Cloud-free}} 
  \startdata
  \multicolumn{11}{c}{Retrieved} \\\hline
  H$_2$O        & $-3.67\substack{+0.12\\-0.15}$ & $\cdots$ & $-3.38\substack{+0.07\\-0.10}$ & $-3.37\substack{+0.06\\-0.09}$ & $-3.56\substack{+0.15\\-0.20}$ & $-3.54\substack{+0.13\\-0.17}$ & $-3.59\substack{+0.10\\-0.15}$ & $-3.67\substack{+0.10\\-0.12}$ & $-3.37\substack{+0.15\\-0.20}$ & $-3.35\substack{+0.12\\-0.16}$\\ 
  CO            & $-3.45\substack{+0.20\\-0.22}$ & $\cdots$ & $-2.92\substack{+0.09\\-0.12}$ & $-2.92\pm0.12$                 & $-3.15\substack{+0.16\\-0.17}$ & $-3.16\pm0.21$                 & $-3.38\substack{+0.17\\-0.20}$ & $-3.42\substack{+0.20\\-0.21}$ & $-2.79\substack{+0.18\\-0.19}$ & $-2.70\substack{+0.20\\-0.18}$\\
  VO            & $-9.31\substack{+0.26\\-0.33}$ & $\cdots$ & $-9.57\substack{+0.53\\-1.43}$ & $-9.19\substack{+0.30\\-1.03}$ & $-9.17\substack{+0.31\\-0.69}$ & $-8.93\substack{+0.22\\-0.27}$ & $-8.86\substack{+0.19\\-0.23}$ & $-8.81\substack{+0.16\\-0.19}$ & $-8.63\substack{+0.28\\-0.40}$ & $-8.33\substack{+0.21\\-0.24}$\\
  CrH           & $-8.41\substack{+0.18\\-0.17}$ & $\cdots$ & $-8.38\substack{+0.23\\-0.29}$ & $-8.13\substack{+0.19\\-0.22}$ & $-8.16\substack{+0.26\\-0.20}$ & $-8.00\substack{+0.20\\-0.22}$ & $-8.00\pm0.15$                 & $-7.95\pm0.16$                 & $-7.70\substack{+0.24\\-0.28}$ & $-7.43\substack{+0.23\\-0.28}$\\
  FeH           & $-8.30\substack{+0.16\\-0.18}$ & $\cdots$ & $-8.40\substack{+0.14\\-0.18}$ & $-8.83\substack{+0.23\\-1.18}$ & $-8.16\substack{+0.16\\-0.19}$ & $-8.40\substack{+0.15\\-0.24}$ & $-8.11\substack{+0.12\\-0.15}$ & $-8.36\substack{+0.12\\-0.18}$ & $-7.96\substack{+0.19\\-0.28}$ & $-8.30\substack{+0.19\\-0.26}$\\
  Na+K          & $-6.34\substack{+0.13\\-0.21}$ & $\cdots$ & $-5.63\substack{+0.18\\-0.24}$ & $-5.43\pm0.19$                 & $-5.86\substack{+0.19\\-0.25}$ & $-5.67\substack{+0.20\\-0.21}$ & $-6.08\substack{+0.17\\-0.25}$ & $-5.96\substack{+0.20\\-0.31}$ & $-5.64\substack{+0.34\\-0.85}$ & $-5.20\substack{+0.31\\-0.58}$ \\
  log $g$ (dex) & $5.25\substack{+0.27\\-0.35}$  & $\cdots$ & $5.50\substack{+0.13\\-0.23}$  & $5.55\substack{+0.07\\-0.18}$  & $5.20\substack{+0.29\\-0.36}$  & $5.27\substack{+0.26\\-0.35}$  & $5.19\substack{+0.21\\-0.28}$  & $5.17\substack{+0.17\\-0.29}$  & $5.20\substack{+0.26\\-0.34}$  & $5.31\substack{+0.13\\-0.25}$ \\ \hline
  \multicolumn{9}{c}{Derived}  \\ \hline
  $L_\mathrm{bol}$          & $-4.22\pm0.01$                   & $\cdots$ & $-4.34\pm 0.01$                   & $-4.35\pm 0.01$                    & $-4.21\pm{0.01}$                 & $-4.21\pm{0.01}$                   & $-3.94\pm 0.01$                   & $-3.94\pm 0.01$                    & $-4.05\pm 0.01$ & $-4.04\pm 0.01$ \\
  $T_\mathrm{eff}$ (K)      &$1892.31\substack{+34.48\\-39.26}$& $\cdots$ & $1719.64\substack{+23.42\\-21.22}$& $1708.33\substack{+20.12\\-19.82}$ &$1859.57\substack{+36.00\\-33.52}$& $1807.83\substack{+27.57\\-25.18}$ & $1926.13\substack{+29.87\\-32.44}$& $1853.87\substack{+27.86\\-24.38}$ & $1920.49\substack{+27.88\\-27.93}$ & $1842.78\substack{+27.16\\-27.55}$ \\
  Radius ($R_\mathrm{Jup}$) & $0.70\substack{+0.04\\-0.03}$    & $\cdots$ & $0.74 \pm 0.06$                   & $0.74 \pm 0.06$                    & $0.74 \pm 0.04$                  & $0.78 \pm 0.03$                    & $ 0.93\pm 0.04$                   & $1.01\pm0.04$.                     & $0.83\pm0.04$ & $0.91\pm0.04$ \\
  Mass ($M_\mathrm{Jup}$)   & $36.36\substack{+28.40\\-19.00}$ & $\cdots$ & $66.80\substack{+24.49\\-25.40}$  & $76.57\substack{+20.21\\ -25.24}$  & $34.15\substack{+30.74\\-18.19}$ & $45.08\substack{+37.60\\-23.87}$   & $53.98\substack{+31.69\\-24.79}$  & $60.62\substack{+26.99\\-28.73}$   & $43.96\substack{+34.62\\-23.24}$ & $68.32\substack{+22.59\\-29.73}$\\
  {C/O}\tablenotemark{a}    & $0.62\substack{+0.08\\-0.10}$    & $\cdots$ & $0.75\substack{+0.03\\-0.04}$     & $0.74\pm0.04$                      & $0.71\substack{+0.07\\-0.05}$    & $0.71\pm0.06$                      & $0.63\substack{+0.08 \\ -0.10}$   & $0.64\substack{+0.08 \\ -0.09}$    & $0.80\pm0.05$  &$0.82\substack{+0.04\\-0.05}$ \\
  $[M/H]$\tablenotemark{a}  & $-0.16\substack{+0.17\\-0.19}$   & $\cdots$ & $0.33\substack{+0.09\\-0.11}$     & $0.34\substack{+0.12 \\ -0.11}$    & $0.11\substack{+0.16\\-0.15}$    & $0.11\substack{+0.20 \\ -0.18}$    & $-0.08\substack{+0.14 \\ -0.17}$  & $-0.14\pm0.17$                     & $0.44\pm0.18$ & $0.53\substack{+0.19\\-0.17}$\\
  \enddata
  \tablenotetext{a}{Atmospheric value.}
  \tablecomments{Cloud models are power-law deck cloud. The best fit model for J1526$+$2043 is the cloud-free model. The cloudy model is shown for comparison. For J0539$-$0059 and J1539$-$0502 the cloudy and cloud-free models are indistinguishable. The best fit model for GD 165B is the cloudy model, with the cloud-free only shown for comparative purposes. Both the cloudy and cloud-free models have uniform-with-altitude gas abundances. Molecular abundances are fractions listed as log values.}
\end{deluxetable*}

Comparing the temperature comparison objects, J0539$-$0059, J1539$-$0520, and GD 165B to J1416+13148A, we find that the C/O ratio and metallicity are consistent with the exception of GD 165B. As well, a majority of the retrieved and derived parameters agree with the exception of the alkalies for J0539$-$0059, CO and VO for GD 165B, and CrH, \Lbol, and the radius for both J1539$-$0520 and GD 165B. Of note, the radius we derive for J1539$-$0520 is much larger than the other objects in this sample. 

When comparing the retrieval-based values to the SED-based values (see Table~\ref{tab:Litdata1416comp}), the \Teff of all comparison objects are $\sim 100-200$~K warmer than the semi-empirical \Teff following the trend seen for J1416+13148A in \cite{Gonz20} and this work, as well as for the unusually read L-dwarf 2MASSW J2224438-015852 (hereafter J2224$-$0158) in \cite{Burn21}. For J1416+13148A, J1526$+$2043, J0539$-$0059, and GD 165B the retrieved radius is smaller than expected from the evolutionary models, which follows the trend seen in previous L dwarf retrieval works (e.g \citealt{Burn17,Gonz20,Burn21}). This is also seen in observational AKARI spectra in \cite{Sora13}, which used a similar method to ours to estimate radii. For most of the mid- to late- L dwarfs in their sample they find smaller radii than predicted from evolutionary models in the range of $0.64-0.81\, R_\mathrm{Jup}$, which encompasses the radii we find for J1416+13148A, J1526$+$2043, J0539$-$0059, and GD 165B. Unlike J1416+13148A and the other comparative objects, the radius values for J1539$-$0520 agree between the two methods. It is unclear why this object is the only one where these two methods agree.

\section{Discussion}\label{sec:discussion} 
\subsection{Comparing P-T Profiles for Objects of the Same Spectral Type}\label{sec:SpTComp}
\begin{figure*}
\gridline{\fig{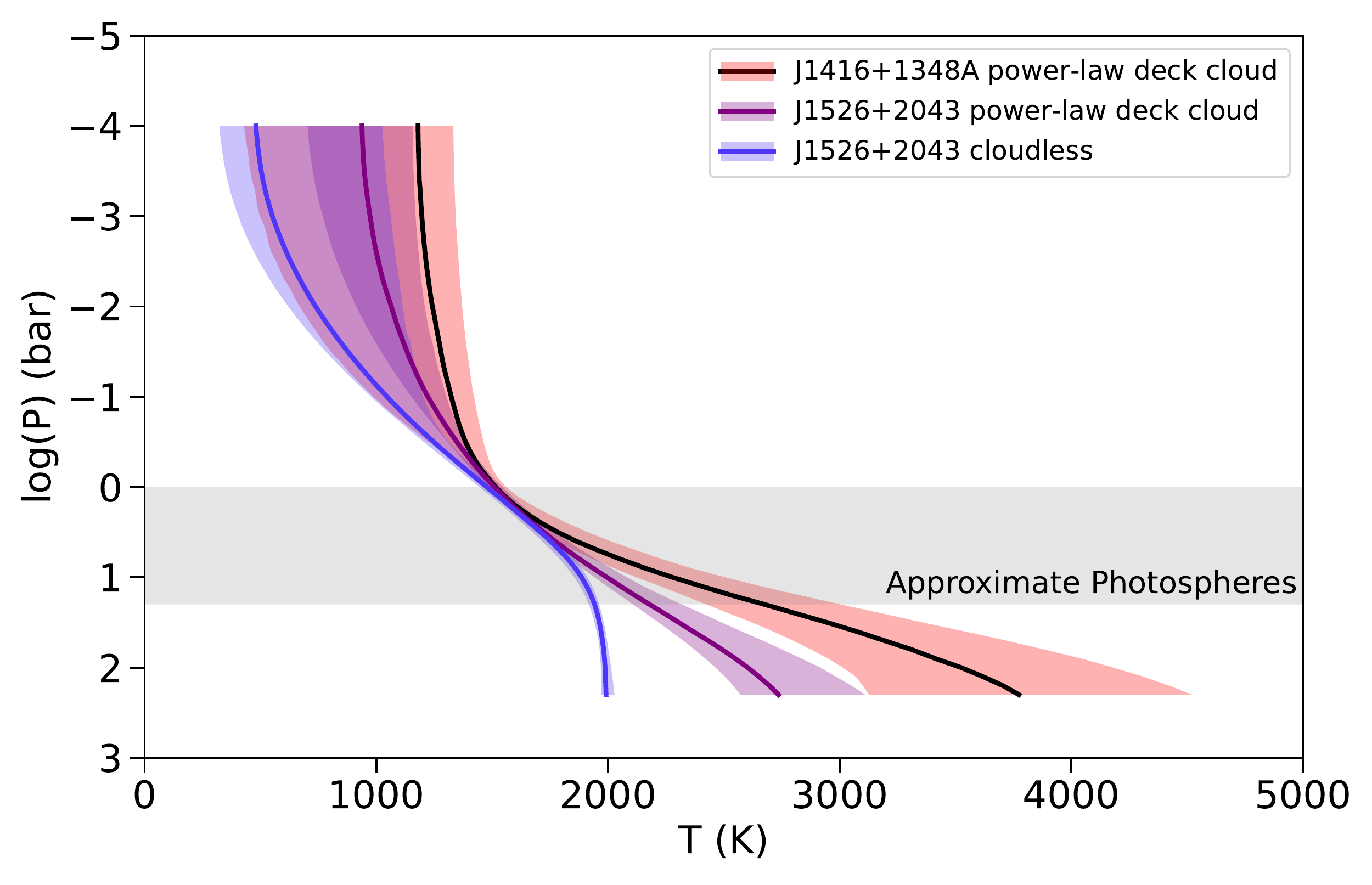}{0.5\textwidth}{\large(a)}}
\gridline{\fig{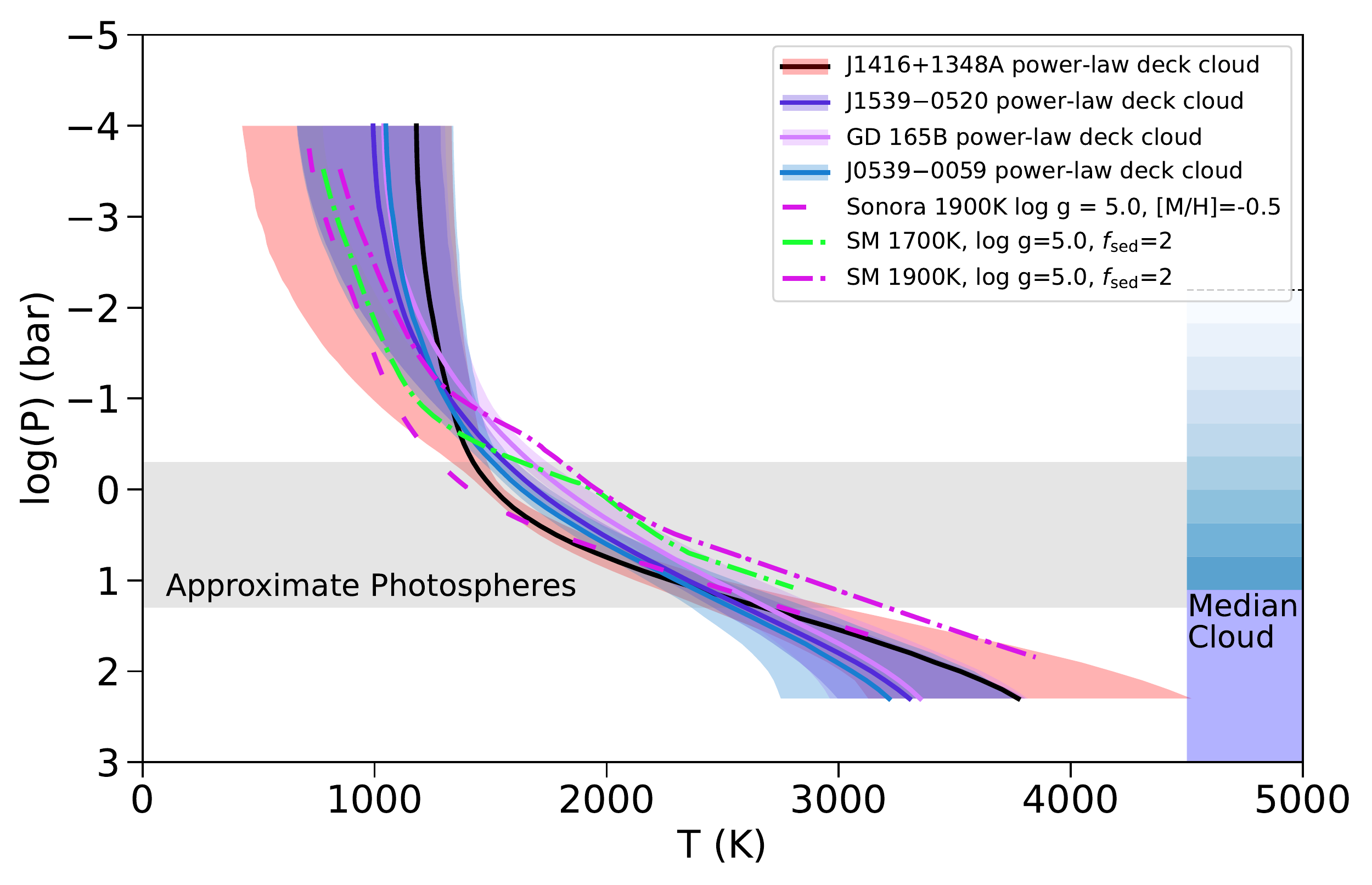}{0.5\textwidth}{\large(b)}
         \fig{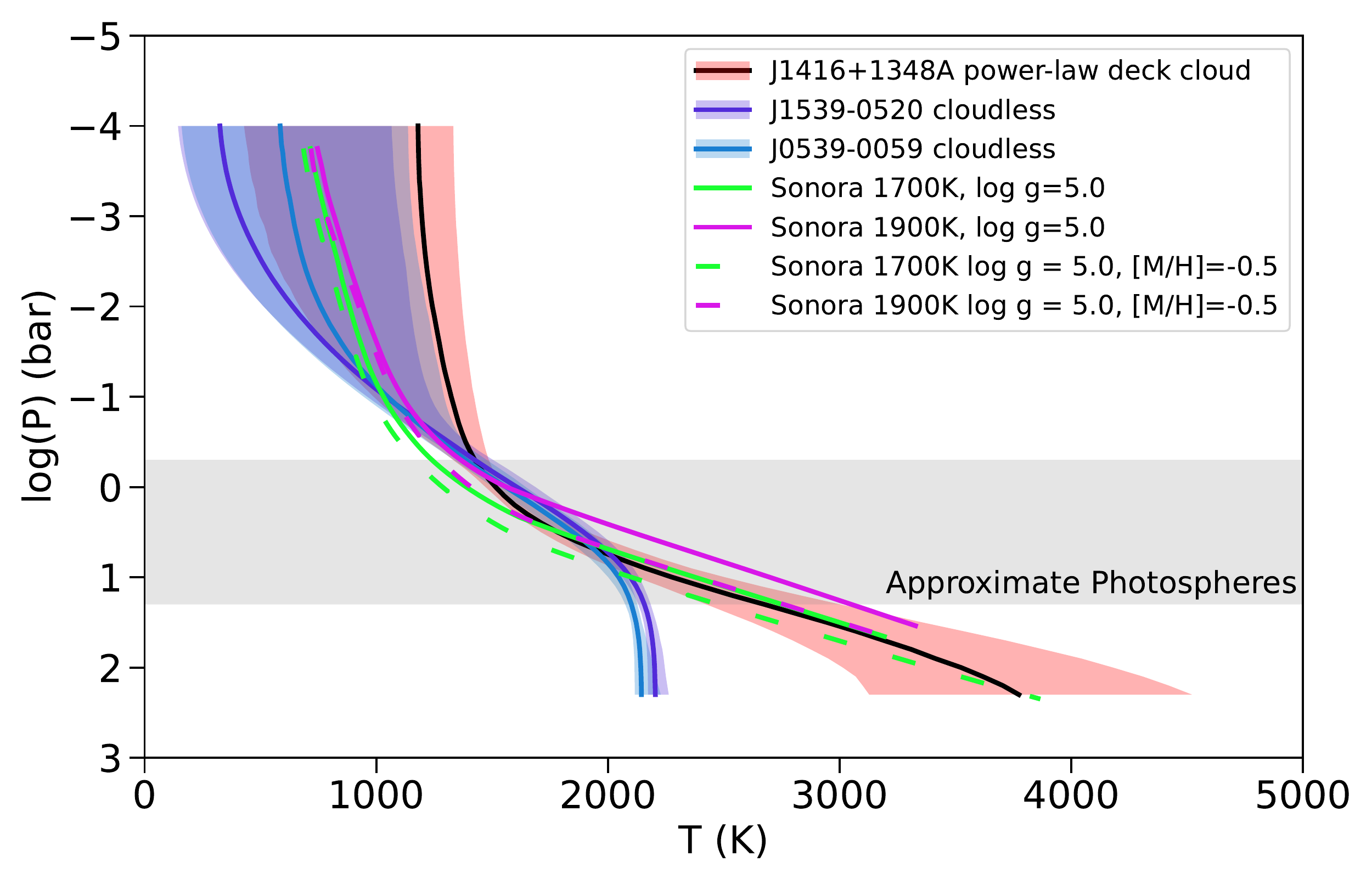}{0.5\textwidth}{\large(c)}} 
\caption{Retrieved Pressure-Temperature Profiles of the sample. For each P-T profile the solid line indicates the median while the shaded region indicates the 1$\sigma$ confidence interval. Grey bar indicates approximate range of the photosphere for the comparison sample. (a) Spectral Type Comparison P-T Profiles: J1526+2043 (purples) and J1416+1348A (black and red). While the cloudless model is the best fit for J1526+2043 we show the power-law deck cloud model for comparison. (b) \Teff Comparison Cloudy Models: J1539$-$0502 (purple), GD 165B (pink), and J0539$-$0059 (blue) compared to J1416+1348A in black and red. Overplotted are the Sonora low-metallicity, log$\,$g$=5.0$ (dotted) model profile \citep{Sonora} and the \cite{Saum08} $f_\mathrm{sed}=2$, log$\,$g$=5.0$ (dot dashed) model profiles which bracket the SED-derived and retrieval-derived effective temperatures (neon green and purple). Far right shows the location of the median cloud top (in purple, where the cloud is optically thick at $\tau=1$) and vertical distribution of the cloud (in the blue shading, showing up to dashed line where the opacity drops to $\tau=0.5$) for J1416+1348A as a guide. Clouds for all comparative objects are in a similar location. (c) \Teff Comparison Cloudless Models: J1539$-$0520 (purple) and J0539$-$0059 (blue) compared to J1416+1348A in black and red. Overplotted are the Sonora solar (solid) and low-metallicity (dashed) model profiles which bracket the SED-derived and retrieval-derived effective temperatures (neon green and purple).}
\label{fig:PT_profile_comparisons}
\vspace{0.5cm} 
\end{figure*}

Figure \ref{fig:PT_profile_comparisons}(a) compares the P-T profiles for the power-law deck cloud models of J1416$+$1348A and J1526$+$2043, as well as the cloudless winning model for J1526$+$2043. The power-law deck cloud was not the winning model for J1526$+$2043 ($\Delta$BIC$=17$), however to compare it to J1416+1348A we show the same cloud model. An important finding is that the retrieved and derived parameters for J1526+2043 agree within 1~$\sigma$ between the cloud-free and power-law deck cloud models (see Table ~\ref{tab:RetrievalParams}). The cloud-free and power-law deck cloud profiles of J1526$+$2043 agree within 1$\sigma$ until they reach pressures of $\sim10$ bar. At this point for the cloudy model, the cloud opacity becomes optically thick and thus we gain no more information below the cloud top for the P-T profile. This is because the shape of the deeper profile is an extrapolation of the profile at lower pressures. For the cloud-free model the turning of the profile to nearly isothermal around 10 bars attempts to recreate the affect of a cloud when one is not present in the model. Therefore because of these affects the retrieved parameters in Table ~\ref{tab:RetrievalParams} are consistent between these two models.

At pressures below $\sim 1$ bar (higher up in the atmosphere) the profiles for J1416+1348A and J1526+2043 agree within 1 $\sigma$. However, at $\sim 1$bar and above (lower down in the atmosphere) J1526+2043 is cooler than J1416+1348A and cools quicker with a steeper profile. While of similar spectral types, J1526+2043 is nearly 180 K cooler than J1416+1348A. This agrees with previous observational findings for L subdwarfs to be $100-400$~K warmer than similar-typed field age L dwarfs \citep{Zhang2017a, Gonz18}. P-T profile comparisons are likely better suited to objects of similar \Teff.

\subsection{Comparing P-T Profiles for Object of the same Temperature} \label{sec:PTcompTeff}
The ``winning'' P-T profiles of the \Teff sample are compared with J1416+1348A in Figures 2(b) and (c) respectively. GD 165B was best fit by a power-law deck cloud model, while J1539$-$0502 and J0539$-$0059 were best fit by a cloudless model and power-law deck cloud model equally (see Table \ref{tab:RetModelscomps} for $\Delta$BIC values). The power-law deck cloud models of J1539$-$0502 and J0539$-$0059, however, have the higher maximum likelihood compared to the cloudless models.

The P-T profiles of \Teff objects generally agree overall to J1416+1348A, much better than the spectral type comparison did (see Figure~\ref{fig:PT_profile_comparisons}(a)). At pressures below $\sim0.1$~bar, J1539$-$0520, GD 165B, and J0539$-$0059 agree within 1$\sigma$ to J1416+1348A. For the cloudy and cloudless models in Figure~\ref{fig:PT_profile_comparisons}(b) and (c), between $\sim0.5-10$~bars the comparative objects are hotter than J1416+1348A. The comparative objects become cooler than J1416+1348A at pressures greater than $\sim 10$ bars.

\begin{figure} 
 \centering
  \includegraphics[scale=0.33]{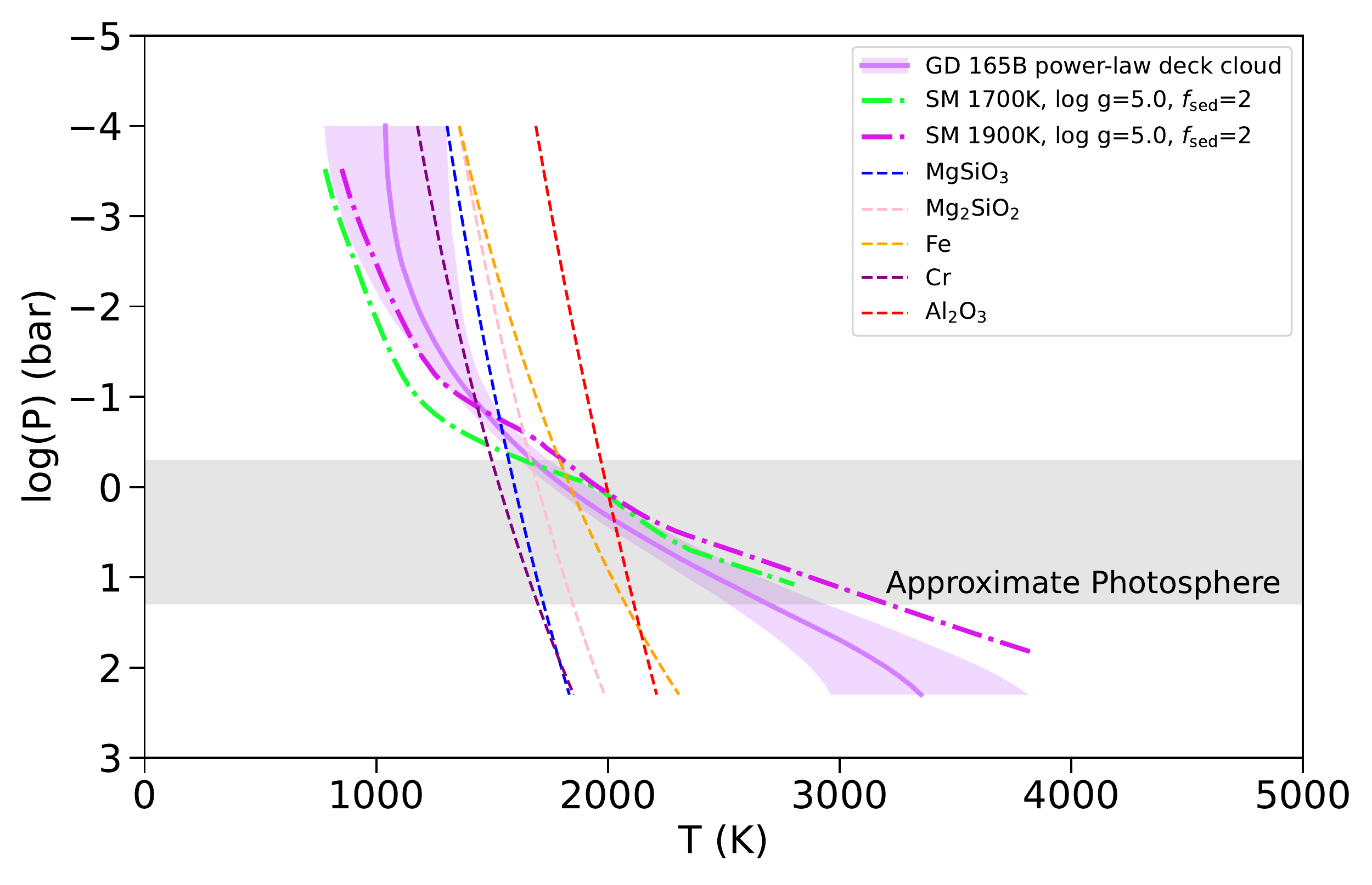}
\caption{Retrieved Pressure-Temperature Profile of GD 165B. Grey bar indicates approximate range of the photosphere. Overplotted are the \cite{Saum08} model profiles that bracket the SED- and retrieval-based \Teff (magenta and neon green dot dashed), along with various cloud species condensation curves (colored dashed).}
\label{fig:GD165B_PT_profile}
\end{figure}

Overplotted in Figure~\ref{fig:PT_profile_comparisons}(b) are the 1900~K, [M/H]$=-0.5$, log$\,g=5.0$ Sonora \citep{Sonora} profile (dotted line) and the \cite{Saum08} (SM) $f_\mathrm{sed}=2$ for temperatures bracketing the semi-empirical and retrieved values. At pressures lower than the photosphere (higher altitudes), we see the median profile for all objects are hotter and more isothermal than the Sonora and SM grid model profiles. Within the photosphere, J1416+1348A's profile agrees with the Sonora cloudless profile within the 1$\sigma$ confidence intervals; while for the comparative objects, the profiles are steeper and cooler than expected from deep adiabat of the SM profiles. This indicates a missing source of opacity in our current retrieval models for the comparative sources, likely coming from an additional cloud. Figure~\ref{fig:GD165B_PT_profile} shows the P-T profile of GD 165B plotted along with the SM grid models and possible condensation curves for various species for these temperatures. Just above the photosphere, the P-T profile of GD 165B intersects with the ensatite (MgSiO$_3$) and forsterite (Mg$_2$SiO$_4$) condensation curves-- two possible additional opacity sources. This has also been previously for J2224$-$0158. When initially retrieved in \cite{Burn17}, only the NIR data was included resulting in a power-law deck cloud, the same cloud model we have for the \Teff comparatives. In \cite{Burn21}, when the MIR spectrum was included, which probes higher altitudes than the NIR spectrum, they found silicate clouds introduced additional opacity sources not visible in the NIR data. As our comparative sources are of similar \Teff to J2224$-$0158 we expect these opacity sources to likely play a role in their atmospheres as well. It should be also noted that the Sonora and SM grid models include gas opacity sources not included in our model (e.g. CO$_2$ and CH$_4$ which are expected to be the next two most abundant gas species in our atmosphere but do not have spectral features in the NIR at these temperatures. We do not constrain these gases for all objects in this sample and while they play a role in shaping our P-T profile it is not as strong of an impact as missing cloud opacites. This is best displayed by the fact that the removal of these gases had little impact on the shape of J1416+1348A's profile. 

One may wonder if wavelength coverage, SNR, metallicity, or variability plays a role in the indistinguishable cloud models for J0539$-$0059 and J1539$-$0520 compared to GD 165B which has a clear preference for the cloudy model. This will be discussed below in Sections~\ref{sec:wavelengthcloud}, \ref{sec:metallicity}, and \ref{sec:SNR}.

\subsection{Does wavelength coverage impact degenerate cloud models?}\label{sec:wavelengthcloud}
To break the degeneracy we have between the cloudless and cloudy models for J0539$-$0059 and J1539$-$0520 we need either (1) a more restricted P-T profile or (2) MIR data which provides additional information about clouds at higher altitudes. By including MIR data, we can get a better handle on the pseudo-continuum (e.g. see Figure~\ref{fig:1416compSED}), which will impact the shape of the P-T profile, as well as additional cloud opacity. While the cloud in the atmosphere of J1416$+$1348A impacts a small portion of its $J$-band spectrum (see \cite{Gonz20} Figures 2(b) and 7(b) and contribution function Figure on Zenodo), the missing clouds in the atmospheres of the \Teff comparative sources likely impact a larger portion of the spectrum and thus contribute more to the P-T profile shape necessitating either of the suggested solutions to break the degeneracy.

Changes to the shape of the P-T profile have been seen previously in the study of the unusually red L-dwarf J2224$-$0158 in \cite{Burn17} and \cite{Burn21}. In \cite{Burn17} they examined J2224$-$0158 using NIR data only and found it was best fit by a power-law deck cloud model with the profile in steeper than expected from the \cite{Saum08} grid model profiles deep in the atmosphere. When the MIR data was included in \cite{Burn21}, the retrieved profile became shallower and thus better matched the expected model profile. Therefore, we expect additional data for J0539$-$0059 and J1539$-$0520 to behave in a similar fashion. J0539$-$0059 currently has MIR Spitzer data which was not included in this retrieval study as we aim to compare the same wavelength coverage for all objects, however, we are be exploring its impact as part of a study of objects with broad wavelength coverage (Gonzales et. al in prep). Therefore we stress the connection between the P-T profile shape and the nature of clouds in a substellar atmosphere in the absence of long wavelength coverage. With a lack of broad wavelength coverage, the P-T profile retrieved can be drastically different than the true shape.

\subsection{Does metallicity impact the P-T profile shape?}\label{sec:metallicity}
Metallicity plays a key role in helping shape the pseudo-continuum of a target's spectrum. As shown by the Sonora grid models in Figure~\ref{fig:PT_profile_comparisons}(c), as the metallicity decreases the ``bend'' in the P-T profile, the transition between the radiative and convective zones, moves diagonally to the bottom left (see the region within the approximate photosphere). It should also be noted that a change in temperature also impacts the ``bend''. This metallicity trend is also seen in the comparative objects in Figure~\ref{fig:PT_profile_comparisons}(b) as they have differing metallicities (see Table~\ref{tab:RetrievalParams} for values). While the differing metallicities of the sample impacts the region between $\sim1-10$ bars where the sample profiles do not overlap, this likely does not have a major impact on their mismatch with the gird model P-T profile deep adiabat because the majority of the \Teff sample is consistent with Solar metallicity. Likely instead, the major impact is the missing cloud opacity for all three \Teff comparative objects.

\subsection{How does signal-to-noise impact the ability to distinguish between cloud models and retrieved parameter uncertainties?}\label{sec:SNR}
\begin{deluxetable*}{l c c c c c c c c}
\tabletypesize{\footnotesize} 
\tablecaption{Signal to Noise of the Spectra Used \label{tab:SNR}}
\tablehead{\colhead{Object} & \colhead{Spectral Type }& \colhead{Retrieved \Teff}& \colhead{Top Model} & \colhead{Second Model} &\multicolumn{4}{c}{Signal-to-Noise} \\
           \cmidrule(lr){6-9}
           \colhead{} & \colhead{} & \colhead{} &  \colhead{} & \colhead{} & \colhead{$YJ$} & \colhead{$H$} & \colhead{$J$ peak} & \colhead{$H$ plateau}}
  \startdata
    J1416$+$1348A& d/sdL7 &$1892.31\substack{+34.48 \\ -39.26}$ &power-law deck cloud, uniform & power-law slab cloud, uniform & 341 & 359 & 437 & 423\\
  J1526$+$2043 & L7 &$1708.33\substack{+20.12\\-19.82}$&cloudless, uniform            & grey deck cloud, CE           & 83 & 99 & 95& 111\\
  J0539$-$0059 & L5 &$1859.57\substack{+36.00\\-33.52}$ &cloudless, uniform\tablenotemark{a}            & power-law deck cloud, uniform\tablenotemark{a} & 228 & 231 & 302 & 227\\
  J1539$-$0520 & L4 &$1926.13\substack{+29.87\\-32.44}$&cloudless, uniform\tablenotemark{a}            & power-law deck cloud, uniform\tablenotemark{a} & 176 & 178 & 219& 198\\
  GD 165B      & L4 &$1920.49\substack{+27.88\\-27.93}$&power-law deck cloud, uniform & cloudless, uniform            & 63 & 67 &79  &78\\
  \enddata
  \tablenotetext{a}{Indistinguishable models.}
  \tablecomments{Uniform = uniform-with-altitude mixing, CE = Chemical Equilibrium. Signal-to-noise calculated between the following wavelengths for $YJ$: $1.0-1.35\, \mu$m, $H$: $1.42-1.80\, \mu$m, $J$ peak: $1.26-1.32\, \mu$m, and $H$ plateau:$1.55-1.70\, \mu$m.}
\end{deluxetable*}  

The SpeX spectra of the comparison sample provides a range of SNR values for each spectrum retrieved. Therefore we evaluate the impact SNR may play on indistinguishable cloud models or gas abundance uncertainties. Table~\ref{tab:SNR} lists the SNR for each of the objects in the sample across various regions in their NIR spectrum. Comparisons of each spectrum's SNR in the sample demonstrate whether SNR impacts the cloud model choice as well as how tightly the retrieved gas abundances can be constrained.

In particular, J1539$-$0520 and GD 165B have the same \Teff and spectral types, but differ in SNR (J1539$-$0520 has a higher SNR spectrum). Therefore we can contrast resultant values to determine the impact SNR may have on scientific conclusions. J1539$-$0520 is a variable source currently thought have cloud variability, and thus we need to consider the impact of more than one source of cloud opacity. However, as discussed in Section~\ref{sec:wavelengthcloud}, NIR data alone cannot probe enough of the atmosphere to best constrain cloud properties for J1539$-$0520. For GD 165B while it is best fit by a cloudy model, it too is likely missing additional cloud opacity as the deep adiabat slope of the P-T profile also differs from the expected grid model slope. Another difference in the two sources is their metallicity, however, as discussed in Section~\ref{sec:metallicity} above, metallicity should impact the relative location of the ``bend'' in the P-T profile (the transition from the adiabatic region to the radiative region). From looking at Table~\ref{tab:SNR}, we can see that SNR does not play a role in constraining the difference between cloudy and cloudless models. Both J1416$+$1348A and GD 165B have distinguishable cloud models with SNR values of 437 and 79 respectively at the $J$-band peak, while J1539$-$0520 has a SNR of 219 at the $J$-band peak and can not differ between cloudy and cloud-free models. It should be noted that these SNRs are relatively high, placing us in a regime where model uncertainties to dominate as opposed to data uncertainties. This may not be the case when examining cloud model differences between spectra with SNR$\lesssim50$. Therefore, the impact of wavelength coverage plays a more significant role in distinguishing between cloud models than the SNR does when SNR$\gtrsim 80$. As clouds affect the spectrum over a wide range of wavelengths there should be no specific features which would require a higher SNR than presented here to be resolved.

One may also ask how SNR impacts the uncertainties for our retrieved parameters. Comparing the retrieved parameters of J1539$-$0520 and GD 165B, the uncertainties for GD 165B are only slightly larger than those of J1539$-$0520. Therefore, SNR could play a minor role in the size of the retrieved parameter uncertainties, however, with a SNR greater than 79 in the $J$-band peak there appears to be no significant difference in the size of the uncertainties.

\subsection{The Impact of Unconstrained Gases on Model Selection}
 When including all of the gases initially used in \cite{Gonz20} for the comparative sample, we found that the ``winning'' model used the thermochemical equilibrium method for J1526+2043 (grey deck cloud) and J0539$-$0059 (power-law deck cloud). However, once the unconstrained gases were removed the ``winning'' models for both J1526+2043 (cloud-free) and J0539$-$0059 (indistinguishable between cloudless and power-law deck cloud) used the uniform-with-altitude method, meaning a free retrieval is preferred. By removing the unconstrained gases from our model, we reduced the number of parameters in the uniform-with-altitude retrieval models. This removed ``excess'' model parameters (those which can't be constrained based on our data) which impact the $\Delta$BIC for the uniform-with-altitude models, but not the thermochemical models. While the uniform-with-altitude method is not physically motivated, the flexibility provided by determining individual gas abundances provides a better fit to the data than the thermochemical equilibrium method which retrieves a C/O ratio and a metallicity instead, and thus is the preferred method once unconstrained parameters are removed. \cite{Burn21}, posed that the challenge in accurately modeling the broadened alkali wings could be a physical reason for this preference. Like seen in their work we too are unable to simultaneously fit the pseudo-continuum near $1\mu$m impacted by the pressure broaden \ion{K}{1} $0.77\mu$m wings and the narrow $J$-band \ion{K}{1} lines. Therefore, the determination of which gases should be included in the retrieval model via iteration is critical for model selection, particularly when deciding between the uniform-with-altitude and thermochemical equilibrium methods.

\section{Conclusions}
Here we present an updated retrieval of J1416$+$1348A and the first retrievals of J1526+2043, J0539$-$0059, J1539$-$0520, and GD 165B to address the question, `` How do the P-T profiles of objects of the same spectral type or \Teff compare to one another?''. From our retrieval models we find that the P-T profiles of objects of similar \Teff look more similar to one another as opposed to objects of the same spectral type. This is critical for subdwarfs as objects of similar spectral type can differ by up to $400$~K. We also noted that a mismatch in the slope at $\sim 0.5$~bar and deeper (adiabatic region)  of our retrieved P-T profiles for J0539$-$0059, J1539$-$0520, and GD 165B compared to the expected shape from the \cite{Saum08} grid models. This indicates that our retrieval models are missing an opacity source, which is likely due to an additional cloud. Retrieval model degeneracies will be even more critical for directly imaged exoplanets due to their limited wavelength coverage. To address this, particularly for objects in which they are unable to distinguish between cloudless and cloudy models, we can either (1) use a more restricted P-T profile on our current data, or (2) include MIR wavelength data as it will allow us to better constrain the pseudo-continuum of our spectrum as well as provide information about cloud opacities at higher altitudes in the atmosphere. Additionally, we noted that the metallicity of a source does not play a large role in the shape of the P-T profile retrieved, but instead missing cloud opacites having a larger impact. We also note that when the input spectrum SNR is $\gtrsim80$, it does not impact the ability to distinguish between various cloud models, but could play a minor role in how tightly the retrieved parameters are constrained. Lastly, we highlight the importance of identifying the ``minimum complexity'' retrieval model, as the addition of unconstrained gases can impact the preference between a uniform-with-altitude gas abundance method and a thermochemical equilibrium gas abundance method.

\acknowledgments 
We thank the anonymous referee for their helpful comment. E.G. thanks Samantha Trumbo for her useful comments. E.G. acknowledges support from the Heising-Simons Foundation for this research. This research was made possible thanks to the Royal Society International Exchange grant No. IES{\textbackslash}R3\textbackslash170266. This research was partially supported by the NSF under Grant No. AST-1614527, grant No. AST-1313278, and grant No. AST-1909776. This publication makes use of data products from the Two Micron All Sky Survey, which is a joint project of the University of Massachusetts and the Infrared Processing and Analysis Center/California Institute of Technology, funded by the National Aeronautics and Space Administration and the National Science Foundation. This publication makes use of data products from the Wide-field Infrared Survey Explorer, which is a joint project of the University of California, Los Angeles, and the Jet Propulsion Laboratory/California Institute of Technology, funded by the National Aeronautics and Space Administration. This work has made use of data from the European Space Agency (ESA) mission {\it Gaia} (\url{https://www.cosmos.esa.int/gaia}), processed by the {\it Gaia} Data Processing and Analysis Consortium (DPAC, \url{https://www.cosmos.esa.int/web/gaia/dpac/consortium}). Funding for the DPAC has been provided by national institutions, in particular the institutions participating in the {\it Gaia} Multilateral Agreement.

\software{astropy \citep{Astropy},  
          SEDkit (\url{https://github.com/hover2pi/SEDkit}, \textit{Eileen Branch}), 
          \textit{Brewster} \citep{Burn17},
          EMCEE \citep{emcee},
          Corner \citep{Corner}
          }

\clearpage
\appendix

\section{Data Used For Generating SEDs}\label{sec:Appendix}
The following tables list all the photometry and spectral information used to generate the SEDs of the samples.

\begin{deluxetable*}{l c c c c c c cc}[h]
\tabletypesize{\normalsize} 
\tablecaption{Photometry used for Comparative sample SEDs \label{tab:photometrySED}}
\tablehead{\colhead{Band} &  \colhead{J1526$+$2043} &\colhead{Ref.} &\colhead{J0539$-$0059} & \colhead{Ref.} & \colhead{J1539$-$0520} & \colhead{Ref.}& \colhead{GD 165B} & \colhead{Ref.}}
  \startdata
  Pan-STARRS $r$       & $\cdots$          & $\cdots$ & $21.076 \pm 0.05$    &  1       & $20.906 \pm 0.06$   & 1        & $\cdots$          & $\cdots$\\
  Pan-STARRS $i$       & $20.301 \pm 0.06$ &      1   & $18.711 \pm 0.11$    &  1       & $18.391 \pm 0.01$   & 1        & $\cdots$          & $\cdots$\\
  Pan-STARRS $z$       & $18.061 \pm 0.01$ &      1   & $16.651 \pm 0.01$    & 1        & $16.541 \pm 0.01$   & 1        & $ 18.87\pm0.01$   & 1\\
  Pan-STARRS $y$       & $17.039 \pm 0.01$ &      1   & $15.559 \pm 0.01$    & 1        & $15.439 \pm 0.01$   & 1        & $\cdots$          & $\cdots$\\
  SDSS $g$             & $\cdots$          & $\cdots$ & $23.49 \pm 0.27$     & 2        & $\cdots$            & $\cdots$ & $22.76\pm0.23$    & 2\\
  SDSS $r$             & $22.81 \pm 0.23$  &      2   & $21.236 \pm 0.061$   & 2        & $\cdots$            & $\cdots$ & $21.255\pm0.073$  & 2\\
  SDSS $i$             & $20.256 \pm 0.05$ &      2   & $18.713 \pm 0.015$   & 2        & $\cdots$            & $\cdots$ & $20.532\pm0.055$  & 2\\
  SDSS $z$             & $17.635 \pm 0.025$&      2   & $16.1929 \pm 0.0093$ & 2        & $\cdots$            & $\cdots$ & $18.456\pm0.031$  & 2\\
  Cousins $R$          & $\cdots$          &  $\cdots$& $\cdots$             & $\cdots$& $19.6882 \pm 0.035$ & 3        & $\cdots$          & $\cdots$\\
  Cousins $I$          & $\cdots$          &  $\cdots$& $\cdots$             & $\cdots$& $17.5614 \pm 0.046$ & 3        & $\cdots$          & $\cdots$\\
  2MASS $J$            & $15.586 \pm 0.055$&  4       & $14.033 \pm 0.031$   &  4       & $13.922 \pm 0.029$ & 4        & $15.687\pm0.078$  &  4\\
  2MASS $H$            & $14.497 \pm 0.044$&  4       & $13.104 \pm 0.026$   &  4       & $13.06 \pm 0.026$  & 4        & $14.781\pm0.07$   &  4 \\
  2MASS $K_\mathrm{s}$ & $13.922 \pm 0.052$&  4       & $12.527 \pm 0.024$   &  4       & $12.575 \pm 0.029$ & 4        & $14.169\pm0.095$  &  4\\ 
  MKO $Y$              &  $\cdots$         & $\cdots$ & $\cdots$             & $\cdots$ & $\cdots$           & $\cdots$ &  $17.01\pm0.1$    & 5 \\
  MKO $J$              &  $\cdots$         & $\cdots$ & $13.85 \pm 0.03$     &  6       & $\cdots$           & $\cdots$ & $15.64\pm0.05$    & 7 \\    
  MKO $H$              &  $\cdots$         & $\cdots$ & $13.04 \pm 0.03$     &  6       & $\cdots$           & $\cdots$ & $14.75\pm0.05$    & 7\\
  MKO $K$              &  $\cdots$         & $\cdots$ & $12.4 \pm 0.03$      &  6       & $\cdots$           & $\cdots$ & $14.09\pm0.05$    & 7\\
  MKO $L^\prime$       &  $\cdots$         & $\cdots$ & $11.32 \pm 0.05$     &  7       & $\cdots$           & $\cdots$ & $12.93\pm0.07$    & 8 \\ 
  MKO $M^\prime$       &  $\cdots$         & $\cdots$ & $11.87 \pm 0.11$     &  9       & $\cdots$           & $\cdots$ & $\cdots$          & $\cdots$\\
  WISE $W1$            & $13.153 \pm 0.025$& 10       & $11.869 \pm 0.023$   & 10       & $12.004 \pm 0.023$ & 10       & $13.198\pm0.024$  & 10 \\
  WISE $W2$            &$12.826 \pm 0.025$ & 10       & $11.578 \pm 0.021$   & 10       & $11.744 \pm 0.022$ & 10       & $13.042\pm0.026$  & 10\\
  WISE $W3$            & $12.54 \pm 0.267$ & 10       &  $\cdots$            & $\cdots$ &$11.655 \pm 0.23$   & 10       & $\cdots$          & $\cdots$\\
  IRAC [3.6 $\mu$m]    & $12.79 \pm 0.02$  & 11       & $11.49 \pm 0.02$     & 11       & $11.65 \pm 0.02$   & 11       & $\cdots$          & $\cdots$\\
  IRAC [4.5 $\mu$m]    & $12.87 \pm 0.03$  & 11       & $11.6 \pm 0.02$      & 11       & $11.75 \pm 0.04$   & 11       & $\cdots$          & $\cdots$\\
  IRAC [5.8 $\mu$m]    & $12.6 \pm 0.11$   & 11       & $11.35 \pm 0.03$     & 11       & $11.61 \pm 0.05$   & 11       & $\cdots$          & $\cdots$\\
  IRAC [8.0 $\mu$m]    & $12.32 \pm 0.04$  & 11       & $11.2 \pm 0.04$      & 11       & $11.6 \pm 0.05$    & 11       & $\cdots$          & $\cdots$\\
  \enddata
\tablerefs{(1) \cite{Cham16}, (2) \cite{Ahn_12}, (3) \cite{Diet14}, (4) \cite{Cutr03}, (5) \cite{Bard14}, (6) \cite{Legg00}, (7) \cite{Legg02a}, (8) \cite{Jone96}, (9) \cite{Goli04a}, (10) \cite{Cutr12}, (11) \cite{Patt06}}
\end{deluxetable*}

\begin{deluxetable*}{l c c c c c c c c c}
\tabletypesize{\small} 
\tablecaption{Spectra used to construct SEDs \label{tab:SpectraSED}}
\tablehead{\colhead{Name} & \colhead{OPT} & \colhead{OPT} & \colhead{OPT} & \colhead{NIR} & \colhead{NIR} & \colhead{NIR} & \colhead{MIR} &\colhead{MIR} & \colhead{MIR} \vspace{-.3cm}\\             \colhead{ } &\colhead{ } &\colhead{Obs. Date} & \colhead{Ref.} &\colhead{ } &\colhead{Obs. Date} & \colhead{Ref.} &\colhead{ } &\colhead{Obs. Date} & \colhead{Ref.}}
  \startdata
  J1526$+$2043 & LRIS & 1998 Sep 20 & 1 & SpeX Prism & 2003 May 23 & 2 & IRS & 2006 Jan 17& 3\\
  J0539$-$0059 & ARC 3.5m: DIS & 1999 Mar 20 & 4 & SpeX SXD\tablenotemark{a} & 2005 Apr 06 & 5 & IRS & 2006 Jan 17& 3\\
  J1539$-$0520 & KPNO 4m: R--C Spec & 2004 Feb 10 & 6 & SpeX Prism & 2008 Jul 14 & 7 & $\cdots$ & $\cdots$ &$\cdots$ \\
  GD 165B      & $\cdots$ & $\cdots$ & $\cdots$ &SpeX Prism & 2009 Jun 29& 7 &$\cdots$ &$\cdots$ &$\cdots$\\\hline
  \enddata
  \tablecomments{For all retrievals, only the NIR SpeX prism spectra from $1-2.5\,\mu$m were used.}
  \tablenotetext{a}{Only SpeX prism spectrum used for retrieval. J0539: 2000 Nov 06, \cite{Schn14}. J1416A: 2009 Jun 28, \cite{Schm10a} }
  \tablenotetext{b}{Combined spectrum of observations on 2011--02--05, 2011--02--12, and 2011--04--13.}
  \tablerefs{(1) \cite{Kirk00}, (2) \cite{Burg04b}, (3) \cite{Cush06b}, (4) \cite{Fan_00}, (5) \cite{Cush05}, (6) \cite{Reid08b}  (7)\cite{Bard14}}
  \tabletypesize{\small} 
\end{deluxetable*}

\clearpage
\bibliographystyle{yahapj}
\bibliography{references}

\begin{thebibliography}{}
\providecommand\natexlab[1]{#1}
\providecommand\JournalTitle[1]{#1}

\bibitem[{{Abazajian} {et~al.}(2009){Abazajian}, {Adelman-McCarthy},
  {Ag{\"u}eros}, {Allam}, {Allende Prieto}, {An}, {Anderson}, {Anderson},
  {Annis}, {Bahcall}, \& et~al.}]{Abaz09}
{Abazajian}, K.~N., {Adelman-McCarthy}, J.~K., {Ag{\"u}eros}, M.~A., {et~al.}
  2009,
  \href{http://dx.doi.org/10.1088/0067-0049/182/2/543}{\JournalTitle{\apjs},
  182, 543}

\bibitem[{{Ahn} {et~al.}(2012){Ahn}, {Alexandroff}, {Allende Prieto},
  {Anderson}, {Anderton}, {Andrews}, {Aubourg}, {Bailey}, {Balbinot}, {Barnes},
  \& et~al.}]{Ahn_12}
{Ahn}, C.~P., {Alexandroff}, R., {Allende Prieto}, C., {et~al.} 2012,
  \href{http://dx.doi.org/10.1088/0067-0049/203/2/21}{\JournalTitle{\apjs},
  203, 21}

\bibitem[{Allers \& Liu(2013)}]{Alle13}
Allers, K.~N., \& Liu, M.~C. 2013,
  \href{http://stacks.iop.org/0004-637X/772/i=2/a=79}{\JournalTitle{The
  Astrophysical Journal}, 772, 79}

\bibitem[{{Allers} {et~al.}(2010){Allers}, {Liu}, {Dupuy}, \&
  {Cushing}}]{Alle10}
{Allers}, K.~N., {Liu}, M.~C., {Dupuy}, T.~J., \& {Cushing}, M.~C. 2010,
  \href{http://dx.doi.org/10.1088/0004-637X/715/1/561}{\JournalTitle{\apj},
  715, 561}

\bibitem[{{Asplund} {et~al.}(2009){Asplund}, {Grevesse}, {Sauval}, \&
  {Scott}}]{Aspl09}
{Asplund}, M., {Grevesse}, N., {Sauval}, A.~J., \& {Scott}, P. 2009,
  \href{http://dx.doi.org/10.1146/annurev.astro.46.060407.145222}{\JournalTitle{\araa},
  47, 481}

\bibitem[{{Astropy Collaboration} {et~al.}(2013){Astropy Collaboration},
  {Robitaille}, {Tollerud}, {Greenfield}, {Droettboom}, {Bray}, {Aldcroft},
  {Davis}, {Ginsburg}, {Price-Whelan}, {Kerzendorf}, {Conley}, {Crighton},
  {Barbary}, {Muna}, {Ferguson}, {Grollier}, {Parikh}, {Nair}, {Unther},
  {Deil}, {Woillez}, {Conseil}, {Kramer}, {Turner}, {Singer}, {Fox}, {Weaver},
  {Zabalza}, {Edwards}, {Azalee Bostroem}, {Burke}, {Casey}, {Crawford},
  {Dencheva}, {Ely}, {Jenness}, {Labrie}, {Lian Lim}, {Pierfederici},
  {Pontzen}, {Ptak}, {Refsdal}, {Servillat}, \& {Streicher}}]{Astropy}
{Astropy Collaboration}, {Robitaille}, T.~P., {Tollerud}, E.~J., {et~al.} 2013,
  \href{http://dx.doi.org/10.1051/0004-6361/201322068}{\JournalTitle{\aap},
  558, A33}

\bibitem[{{Bailer-Jones}(2008)}]{Bail08}
{Bailer-Jones}, C.~A.~L. 2008,
  \href{http://dx.doi.org/10.1111/j.1365-2966.2007.12781.x}{\JournalTitle{\mnras},
  384, 1145}

\bibitem[{{Bailer-Jones} \& {Mundt}(2001)}]{Bail01}
{Bailer-Jones}, C.~A.~L., \& {Mundt}, R. 2001,
  \href{http://dx.doi.org/10.1051/0004-6361:20000416}{\JournalTitle{\aap}, 367,
  218}

\bibitem[{{Baraffe} {et~al.}(2003){Baraffe}, {Chabrier}, {Barman}, {Allard}, \&
  {Hauschildt}}]{Bara03}
{Baraffe}, I., {Chabrier}, G., {Barman}, T.~S., {Allard}, F., \& {Hauschildt},
  P.~H. 2003,
  \href{http://dx.doi.org/10.1051/0004-6361:20030252}{\JournalTitle{\aap}, 402,
  701}

\bibitem[{{Bardalez Gagliuffi} {et~al.}(2014){Bardalez Gagliuffi}, {Burgasser},
  {Gelino}, {Looper}, {Nicholls}, {Schmidt}, {Cruz}, {West}, {Gizis}, \&
  {Metchev}}]{Bard14}
{Bardalez Gagliuffi}, D.~C., {Burgasser}, A.~J., {Gelino}, C.~R., {et~al.}
  2014,
  \href{http://dx.doi.org/10.1088/0004-637X/794/2/143}{\JournalTitle{\apj},
  794, 143}

\bibitem[{{Becklin} \& {Zuckerman}(1988)}]{Beck88}
{Becklin}, E.~E., \& {Zuckerman}, B. 1988,
  \href{http://dx.doi.org/10.1038/336656a0}{\JournalTitle{\nat}, 336, 656}

\bibitem[{{Bell} {et~al.}(2015){Bell}, {Mamajek}, \& {Naylor}}]{Bell15}
{Bell}, C.~P.~M., {Mamajek}, E.~E., \& {Naylor}, T. 2015,
  \href{http://dx.doi.org/10.1093/mnras/stv1981}{\JournalTitle{\mnras}, 454,
  593}

\bibitem[{{Blake} {et~al.}(2010){Blake}, {Charbonneau}, \& {White}}]{Blak10}
{Blake}, C.~H., {Charbonneau}, D., \& {White}, R.~J. 2010,
  \href{http://dx.doi.org/10.1088/0004-637X/723/1/684}{\JournalTitle{\apj},
  723, 684}

\bibitem[{{Bowler} {et~al.}(2010){Bowler}, {Liu}, \& {Dupuy}}]{Bowl10}
{Bowler}, B.~P., {Liu}, M.~C., \& {Dupuy}, T.~J. 2010,
  \href{http://dx.doi.org/10.1088/0004-637X/710/1/45}{\JournalTitle{\apj}, 710,
  45}

\bibitem[{{Buenzli} {et~al.}(2014){Buenzli}, {Apai}, {Radigan}, {Reid}, \&
  {Flateau}}]{Buen14}
{Buenzli}, E., {Apai}, D., {Radigan}, J., {Reid}, I.~N., \& {Flateau}, D. 2014,
  \href{http://dx.doi.org/10.1088/0004-637X/782/2/77}{\JournalTitle{\apj}, 782,
  77}

\bibitem[{{Burgasser} {et~al.}(2004){Burgasser}, {McElwain}, {Kirkpatrick},
  {Cruz}, {Tinney}, \& {Reid}}]{Burg04b}
{Burgasser}, A.~J., {McElwain}, M.~W., {Kirkpatrick}, J.~D., {et~al.} 2004,
  \href{http://dx.doi.org/10.1086/383549}{\JournalTitle{\aj}, 127, 2856}

\bibitem[{{Burgasser} {et~al.}(2008){Burgasser}, {Vrba}, {L{\'e}pine}, {Munn},
  {Luginbuhl}, {Henden}, {Guetter}, \& {Canzian}}]{Burg08a}
{Burgasser}, A.~J., {Vrba}, F.~J., {L{\'e}pine}, S., {et~al.} 2008,
  \href{http://dx.doi.org/10.1086/523810}{\JournalTitle{\apj}, 672, 1159}

\bibitem[{{Burgasser} {et~al.}(2009){Burgasser}, {Witte}, {Helling},
  {Sanderson}, {Bochanski}, \& {Hauschildt}}]{Burg09a}
{Burgasser}, A.~J., {Witte}, S., {Helling}, C., {et~al.} 2009,
  \href{http://dx.doi.org/10.1088/0004-637X/697/1/148}{\JournalTitle{\apj},
  697, 148}

\bibitem[{{Burgasser} {et~al.}(2002){Burgasser}, {Kirkpatrick}, {Brown},
  {Reid}, {Burrows}, {Liebert}, {Matthews}, {Gizis}, {Dahn}, {Monet}, {Cutri},
  \& {Skrutskie}}]{Burg02a}
{Burgasser}, A.~J., {Kirkpatrick}, J.~D., {Brown}, M.~E., {et~al.} 2002,
  \href{http://dx.doi.org/10.1086/324033}{\JournalTitle{\apj}, 564, 421}

\bibitem[{{Burgasser} {et~al.}(2003){Burgasser}, {Kirkpatrick}, {Burrows},
  {Liebert}, {Reid}, {Gizis}, {McGovern}, {Prato}, \& {McLean}}]{Burg03c}
{Burgasser}, A.~J., {Kirkpatrick}, J.~D., {Burrows}, A., {et~al.} 2003,
  \href{http://dx.doi.org/10.1086/375813}{\JournalTitle{\apj}, 592, 1186}

\bibitem[{{Burgasser} {et~al.}(2015){Burgasser}, {Logsdon}, {Gagn{\'e}},
  {Bochanski}, {Faherty}, {West}, {Mamajek}, {Schmidt}, \& {Cruz}}]{Burg15}
{Burgasser}, A.~J., {Logsdon}, S.~E., {Gagn{\'e}}, J., {et~al.} 2015,
  \href{http://dx.doi.org/10.1088/0067-0049/220/1/18}{\JournalTitle{\apjs},
  220, 18}

\bibitem[{{Burningham} {et~al.}(2017){Burningham}, {Marley}, {Line}, {Lupu},
  {Visscher}, {Morley}, {Saumon}, \& {Freedman}}]{Burn17}
{Burningham}, B., {Marley}, M.~S., {Line}, M.~R., {et~al.} 2017,
  \href{http://dx.doi.org/10.1093/mnras/stx1246}{\JournalTitle{\mnras}, 470,
  1177}

\bibitem[{{Burningham} {et~al.}(2010){Burningham}, {Leggett}, {Lucas},
  {Pinfield}, {Smart}, {Day-Jones}, {Jones}, {Murray}, {Nickson}, {Tamura},
  {Zhang}, {Lodieu}, {Tinney}, \& {Zapatero Osorio}}]{Burn10}
{Burningham}, B., {Leggett}, S.~K., {Lucas}, P.~W., {et~al.} 2010,
  \href{http://dx.doi.org/10.1111/j.1365-2966.2010.16411.x}{\JournalTitle{\mnras},
  404, 1952}

\bibitem[{{Burningham} {et~al.}(2021){Burningham}, {Faherty}, {Gonzales},
  {Marley}, {Visscher}, {Lupu}, {Gaarn}, {Fabienne Bieger}, {Freedman}, \&
  {Saumon}}]{Burn21}
{Burningham}, B., {Faherty}, J.~K., {Gonzales}, E.~C., {et~al.} 2021,
  \href{http://dx.doi.org/10.1093/mnras/stab1361}{\JournalTitle{\mnras}, 506,
  1944}

\bibitem[{{Chabrier} \& {Baraffe}(1997)}]{Chab97}
{Chabrier}, G., \& {Baraffe}, I. 1997, \JournalTitle{\aap}, 327, 1039

\bibitem[{{Chabrier} {et~al.}(2000){Chabrier}, {Baraffe}, {Allard}, \&
  {Hauschildt}}]{Chab00}
{Chabrier}, G., {Baraffe}, I., {Allard}, F., \& {Hauschildt}, P. 2000,
  \href{http://dx.doi.org/10.1086/309513}{\JournalTitle{\apj}, 542, 464}

\bibitem[{{Chambers} {et~al.}(2016){Chambers}, {Magnier}, {Metcalfe},
  {Flewelling}, {Huber}, {Waters}, {Denneau}, {Draper}, {Farrow}, {Finkbeiner},
  {Holmberg}, {Koppenhoefer}, {Price}, {Rest}, {Saglia}, {Schlafly}, {Smartt},
  {Sweeney}, {Wainscoat}, {Burgett}, {Chastel}, {Grav}, {Heasley}, {Hodapp},
  {Jedicke}, {Kaiser}, {Kudritzki}, {Luppino}, {Lupton}, {Monet}, {Morgan},
  {Onaka}, {Shiao}, {Stubbs}, {Tonry}, {White}, {Ba{\~n}ados}, {Bell},
  {Bender}, {Bernard}, {Boegner}, {Boffi}, {Botticella}, {Calamida},
  {Casertano}, {Chen}, {Chen}, {Cole}, {Deacon}, {Frenk}, {Fitzsimmons},
  {Gezari}, {Gibbs}, {Goessl}, {Goggia}, {Gourgue}, {Goldman}, {Grant},
  {Grebel}, {Hambly}, {Hasinger}, {Heavens}, {Heckman}, {Henderson}, {Henning},
  {Holman}, {Hopp}, {Ip}, {Isani}, {Jackson}, {Keyes}, {Koekemoer}, {Kotak},
  {Le}, {Liska}, {Long}, {Lucey}, {Liu}, {Martin}, {Masci}, {McLean}, {Mindel},
  {Misra}, {Morganson}, {Murphy}, {Obaika}, {Narayan}, {Nieto-Santisteban},
  {Norberg}, {Peacock}, {Pier}, {Postman}, {Primak}, {Rae}, {Rai}, {Riess},
  {Riffeser}, {Rix}, {R{\"o}ser}, {Russel}, {Rutz}, {Schilbach}, {Schultz},
  {Scolnic}, {Strolger}, {Szalay}, {Seitz}, {Small}, {Smith}, {Soderblom},
  {Taylor}, {Thomson}, {Taylor}, {Thakar}, {Thiel}, {Thilker}, {Unger},
  {Urata}, {Valenti}, {Wagner}, {Walder}, {Walter}, {Watters}, {Werner},
  {Wood-Vasey}, \& {Wyse}}]{Cham16}
{Chambers}, K.~C., {Magnier}, E.~A., {Metcalfe}, N., {et~al.} 2016,
  \JournalTitle{arXiv e-prints},
  \href{http://arxiv.org/abs/1612.05560}{{\sffamily arXiv:1612.05560
  [astro-ph.IM]}}

\bibitem[{{Cruz} {et~al.}(2009){Cruz}, {Kirkpatrick}, \& {Burgasser}}]{Cruz09}
{Cruz}, K.~L., {Kirkpatrick}, J.~D., \& {Burgasser}, A.~J. 2009,
  \href{http://dx.doi.org/10.1088/0004-6256/137/2/3345}{\JournalTitle{\aj},
  137, 3345}

\bibitem[{{Cruz} {et~al.}(2018){Cruz}, {N{\'u}{\~n}ez}, {Burgasser},
  {Abrahams}, {Rice}, {Reid}, \& {Looper}}]{Cruz18}
{Cruz}, K.~L., {N{\'u}{\~n}ez}, A., {Burgasser}, A.~J., {et~al.} 2018,
  \href{http://dx.doi.org/10.3847/1538-3881/aa9d8a}{\JournalTitle{\aj}, 155,
  34}

\bibitem[{{Cushing} {et~al.}(2009){Cushing}, {Looper}, {Burgasser},
  {Kirkpatrick}, {Faherty}, {Cruz}, {Sweet}, \& {Sanderson}}]{Cush09}
{Cushing}, M.~C., {Looper}, D., {Burgasser}, A.~J., {et~al.} 2009,
  \href{http://dx.doi.org/10.1088/0004-637X/696/1/986}{\JournalTitle{\apj},
  696, 986}

\bibitem[{{Cushing} {et~al.}(2005){Cushing}, {Rayner}, \& {Vacca}}]{Cush05}
{Cushing}, M.~C., {Rayner}, J.~T., \& {Vacca}, W.~D. 2005,
  \href{http://dx.doi.org/10.1086/428040}{\JournalTitle{\apj}, 623, 1115}

\bibitem[{{Cushing} {et~al.}(2010){Cushing}, {Saumon}, \& {Marley}}]{Cush10}
{Cushing}, M.~C., {Saumon}, D., \& {Marley}, M.~S. 2010,
  \href{http://dx.doi.org/10.1088/0004-6256/140/5/1428}{\JournalTitle{\aj},
  140, 1428}

\bibitem[{{Cushing} {et~al.}(2006){Cushing}, {Roellig}, {Marley}, {Saumon},
  {Leggett}, {Kirkpatrick}, {Wilson}, {Sloan}, {Mainzer}, {Van Cleve}, \&
  {Houck}}]{Cush06b}
{Cushing}, M.~C., {Roellig}, T.~L., {Marley}, M.~S., {et~al.} 2006,
  \href{http://dx.doi.org/10.1086/505637}{\JournalTitle{\apj}, 648, 614}

\bibitem[{{Cushing} {et~al.}(2011){Cushing}, {Kirkpatrick}, {Gelino},
  {Griffith}, {Skrutskie}, {Mainzer}, {Marsh}, {Beichman}, {Burgasser},
  {Prato}, {Simcoe}, {Marley}, {Saumon}, {Freedman}, {Eisenhardt}, \&
  {Wright}}]{Cush11}
{Cushing}, M.~C., {Kirkpatrick}, J.~D., {Gelino}, C.~R., {et~al.} 2011,
  \href{http://dx.doi.org/10.1088/0004-637X/743/1/50}{\JournalTitle{\apj}, 743,
  50}

\bibitem[{{Cutri} \& {et al.}(2012)}]{Cutr12}
{Cutri}, R.~M., \& {et al.} 2012, \JournalTitle{VizieR Online Data Catalog},
  2311

\bibitem[{{Cutri} {et~al.}(2003){Cutri}, {Skrutskie}, {van Dyk}, {Beichman},
  {Carpenter}, {Chester}, {Cambresy}, {Evans}, {Fowler}, {Gizis}, {Howard},
  {Huchra}, {Jarrett}, {Kopan}, {Kirkpatrick}, {Light}, {Marsh}, {McCallon},
  {Schneider}, {Stiening}, {Sykes}, {Weinberg}, {Wheaton}, {Wheelock}, \&
  {Zacarias}}]{Cutr03}
{Cutri}, R.~M., {Skrutskie}, M.~F., {van Dyk}, S., {et~al.} 2003, {2MASS All
  Sky Catalog of point sources.}

\bibitem[{{Dahn} {et~al.}(2008){Dahn}, {Harris}, {Levine}, {Tilleman}, {Monet},
  {Stone}, {Guetter}, {Canzian}, {Pier}, {Hartkopf}, {Liebert}, \&
  {Cushing}}]{Dahn08}
{Dahn}, C.~C., {Harris}, H.~C., {Levine}, S.~E., {et~al.} 2008,
  \href{http://dx.doi.org/10.1086/591050}{\JournalTitle{\apj}, 686, 548}

\bibitem[{{Dieterich} {et~al.}(2014){Dieterich}, {Henry}, {Jao}, {Winters},
  {Hosey}, {Riedel}, \& {Subasavage}}]{Diet14}
{Dieterich}, S.~B., {Henry}, T.~J., {Jao}, W.-C., {et~al.} 2014,
  \href{http://dx.doi.org/10.1088/0004-6256/147/5/94}{\JournalTitle{\aj}, 147,
  94}

\bibitem[{{Dupuy} \& {Liu}(2017)}]{Dupu17}
{Dupuy}, T.~J., \& {Liu}, M.~C. 2017,
  \href{http://dx.doi.org/10.3847/1538-4365/aa5e4c}{\JournalTitle{\apjs}, 231,
  15}

\bibitem[{{Faherty} {et~al.}(2009){Faherty}, {Burgasser}, {Cruz}, {Shara},
  {Walter}, \& {Gelino}}]{Fahe09}
{Faherty}, J.~K., {Burgasser}, A.~J., {Cruz}, K.~L., {et~al.} 2009,
  \href{http://dx.doi.org/10.1088/0004-6256/137/1/1}{\JournalTitle{\aj}, 137,
  1}

\bibitem[{{Faherty} {et~al.}(2012){Faherty}, {Burgasser}, {Walter}, {Van der
  Bliek}, {Shara}, {Cruz}, {West}, {Vrba}, \& {Anglada-Escud{\'e}}}]{Fahe12}
{Faherty}, J.~K., {Burgasser}, A.~J., {Walter}, F.~M., {et~al.} 2012,
  \href{http://dx.doi.org/10.1088/0004-637X/752/1/56}{\JournalTitle{\apj}, 752,
  56}

\bibitem[{{Faherty} {et~al.}(2016){Faherty}, {Riedel}, {Cruz}, {Gagne},
  {Filippazzo}, {Lambrides}, {Fica}, {Weinberger}, {Thorstensen}, {Tinney},
  {Baldassare}, {Lemonier}, \& {Rice}}]{Fahe16}
{Faherty}, J.~K., {Riedel}, A.~R., {Cruz}, K.~L., {et~al.} 2016,
  \href{http://dx.doi.org/10.3847/0067-0049/225/1/10}{\JournalTitle{\apjs},
  225, 10}

\bibitem[{{Fan} {et~al.}(2000){Fan}, {Knapp}, {Strauss}, {Gunn}, {Lupton},
  {Ivezi{\'c}}, {Rockosi}, {Yanny}, {Kent}, {Schneider}, {Kirkpatrick},
  {Annis}, {Bastian}, {Berman}, {Brinkmann}, {Csabai}, {Federwitz}, {Fukugita},
  {Gurbani}, {Hennessy}, {Hindsley}, {Ichikawa}, {Lamb}, {Lindenmeyer},
  {Mantsch}, {McKay}, {Munn}, {Nash}, {Okamura}, {Pauls}, {Pier},
  {Rechenmacher}, {Rivetta}, {Sergey}, {Stoughton}, {Szalay}, {Szokoly},
  {Tucker}, {York}, \& {SDSS Collaboration}}]{Fan_00}
{Fan}, X., {Knapp}, G.~R., {Strauss}, M.~A., {et~al.} 2000,
  \href{http://dx.doi.org/10.1086/301224}{\JournalTitle{\aj}, 119, 928}

\bibitem[{{Fegley} \& {Lodders}(1994)}]{Fegl94}
{Fegley}, Jr., B., \& {Lodders}, K. 1994,
  \href{http://dx.doi.org/10.1006/icar.1994.1111}{\JournalTitle{Icarus}, 110,
  117}

\bibitem[{{Fegley} \& {Lodders}(1996)}]{Fegl96}
---. 1996, \href{http://dx.doi.org/10.1086/310356}{\JournalTitle{\apjl}, 472,
  L37}

\bibitem[{{Filippazzo} {et~al.}(2015){Filippazzo}, {Rice}, {Faherty}, {Cruz},
  {Van Gordon}, \& {Looper}}]{Fili15}
{Filippazzo}, J.~C., {Rice}, E.~L., {Faherty}, J., {et~al.} 2015,
  \href{http://dx.doi.org/10.1088/0004-637X/810/2/158}{\JournalTitle{\apj},
  810, 158}

\bibitem[{Foreman-Mackey(2016)}]{Corner}
Foreman-Mackey, D. 2016,
  \href{http://dx.doi.org/10.21105/joss.00024}{\JournalTitle{Journal of Open
  Source Software}, 1, 24}

\bibitem[{{Foreman-Mackey} {et~al.}(2013){Foreman-Mackey}, {Hogg}, {Lang}, \&
  {Goodman}}]{emcee}
{Foreman-Mackey}, D., {Hogg}, D.~W., {Lang}, D., \& {Goodman}, J. 2013,
  \href{http://dx.doi.org/10.1086/670067}{\JournalTitle{\pasp}, 125, 306}

\bibitem[{{Gagn{\'e}} {et~al.}(2017){Gagn{\'e}}, {Faherty}, {Burgasser},
  {Artigau}, {Bouchard}, {Albert}, {Lafreni{\`e}re}, {Doyon}, \& {Bardalez
  Gagliuffi}}]{Gagn17}
{Gagn{\'e}}, J., {Faherty}, J.~K., {Burgasser}, A.~J., {et~al.} 2017,
  \href{http://dx.doi.org/10.3847/2041-8213/aa70e2}{\JournalTitle{\apjl}, 841,
  L1}

\bibitem[{{Gaia Collaboration} {et~al.}(2018){Gaia Collaboration}, {Brown, A.
  G. A.}, {Vallenari, A.}, {Prusti, T.}, {de Bruijne, J. H. J.}, \& {et
  al.}}]{GaiaDR2}
{Gaia Collaboration}, {Brown, A. G. A.}, {Vallenari, A.}, {et~al.} 2018,
  \href{http://dx.doi.org/10.1051/0004-6361/201833051}{\JournalTitle{A\&A}}

\bibitem[{{Gaia Collaboration} {et~al.}(2016){Gaia Collaboration}, {Brown},
  {Vallenari}, {Prusti}, {de Bruijne}, {Mignard}, {Drimmel}, {Babusiaux},
  {Bailer-Jones}, {Bastian}, \& et~al.}]{GaiaDR1}
{Gaia Collaboration}, {Brown}, A.~G.~A., {Vallenari}, A., {et~al.} 2016,
  \href{http://dx.doi.org/10.1051/0004-6361/201629512}{\JournalTitle{\aap},
  595, A2}

\bibitem[{{Geballe} {et~al.}(2002){Geballe}, {Knapp}, {Leggett}, {Fan},
  {Golimowski}, {Anderson}, {Brinkmann}, {Csabai}, {Gunn}, {Hawley},
  {Hennessy}, {Henry}, {Hill}, {Hindsley}, {Ivezi{\'c}}, {Lupton}, {McDaniel},
  {Munn}, {Narayanan}, {Peng}, {Pier}, {Rockosi}, {Schneider}, {Smith},
  {Strauss}, {Tsvetanov}, {Uomoto}, {York}, \& {Zheng}}]{Geba02}
{Geballe}, T.~R., {Knapp}, G.~R., {Leggett}, S.~K., {et~al.} 2002,
  \href{http://dx.doi.org/10.1086/324078}{\JournalTitle{\apj}, 564, 466}

\bibitem[{{Gharib-Nezhad} {et~al.}(2021){Gharib-Nezhad}, {Iyer}, {Line},
  {Freedman}, {Marley}, \& {Batalha}}]{Ghar21}
{Gharib-Nezhad}, E., {Iyer}, A.~R., {Line}, M.~R., {et~al.} 2021,
  \JournalTitle{arXiv e-prints}, arXiv:2104.00264

\bibitem[{{Gizis}(1997)}]{Gizi97}
{Gizis}, J.~E. 1997,
  \href{http://dx.doi.org/10.1086/118302}{\JournalTitle{\aj}, 113, 806}

\bibitem[{{Golimowski} {et~al.}(2004){Golimowski}, {Leggett}, {Marley}, {Fan},
  {Geballe}, {Knapp}, {Vrba}, {Henden}, {Luginbuhl}, {Guetter}, {Munn},
  {Canzian}, {Zheng}, {Tsvetanov}, {Chiu}, {Glazebrook}, {Hoversten},
  {Schneider}, \& {Brinkmann}}]{Goli04a}
{Golimowski}, D.~A., {Leggett}, S.~K., {Marley}, M.~S., {et~al.} 2004,
  \href{http://dx.doi.org/10.1086/420709}{\JournalTitle{\aj}, 127, 3516}

\bibitem[{{Gonzales} {et~al.}(2020){Gonzales}, {Burningham}, {Faherty},
  {Cleary}, {Visscher}, {Marley}, {Lupu}, \& {Freedman}}]{Gonz20}
{Gonzales}, E.~C., {Burningham}, B., {Faherty}, J.~K., {et~al.} 2020,
  \href{http://dx.doi.org/10.3847/1538-4357/abbee2}{\JournalTitle{\apj}, 905,
  46}

\bibitem[{{Gonzales} {et~al.}(2021){Gonzales}, {Burningham}, {Faherty},
  {Visscher}, {Marley}, {Lupu}, {Freedman}, \& {Lewis}}]{Gonz21}
---. 2021,
  \href{http://dx.doi.org/10.3847/1538-4357/ac294e}{\JournalTitle{\apj}, 923,
  19}

\bibitem[{{Gonzales} {et~al.}(2018){Gonzales}, {Faherty}, {Gagn{\'e}},
  {Artigau}, \& {Bardalez Gagliuffi}}]{Gonz18}
{Gonzales}, E.~C., {Faherty}, J.~K., {Gagn{\'e}}, J., {Artigau}, {\'E}., \&
  {Bardalez Gagliuffi}, D. 2018,
  \href{http://stacks.iop.org/0004-637X/864/i=1/a=100}{\JournalTitle{\apj},
  864, 100}

\bibitem[{{Gonzales} {et~al.}(2019){Gonzales}, {Faherty}, {Gagn{\'e}}, {Teske},
  {McWilliam}, \& {Cruz}}]{Gonz19}
{Gonzales}, E.~C., {Faherty}, J.~K., {Gagn{\'e}}, J., {et~al.} 2019,
  \href{http://dx.doi.org/10.3847/1538-4357/ab48fc}{\JournalTitle{\apj}, 886,
  131}

\bibitem[{{Hayashi} \& {Nakano}(1963)}]{Haya63}
{Hayashi}, C., \& {Nakano}, T. 1963,
  \href{http://dx.doi.org/10.1143/PTP.30.460}{\JournalTitle{Progress of
  Theoretical Physics}, 30, 460}

\bibitem[{{Howe} {et~al.}(2022){Howe}, {McElwain}, \& {Mandell}}]{Howe22}
{Howe}, A.~R., {McElwain}, M.~W., \& {Mandell}, A.~M. 2022, \JournalTitle{arXiv
  e-prints}, arXiv:2203.11706

\bibitem[{{Jones} {et~al.}(1996){Jones}, {Longmore}, {Allard}, \&
  {Hauschildt}}]{Jone96}
{Jones}, H. R.~A., {Longmore}, A.~J., {Allard}, F., \& {Hauschildt}, P.~H.
  1996, \href{http://dx.doi.org/10.1093/mnras/280.1.77}{\JournalTitle{\mnras},
  280, 77}

\bibitem[{{Karalidi} {et~al.}(2021){Karalidi}, {Marley}, {Fortney}, {Morley},
  {Saumon}, {Lupu}, {Visscher}, \& {Freedman}}]{Kara21}
{Karalidi}, T., {Marley}, M., {Fortney}, J.~J., {et~al.} 2021,
  \href{http://dx.doi.org/10.3847/1538-4357/ac3140}{\JournalTitle{\apj}, 923,
  269}

\bibitem[{Kass \& Raftery(1995)}]{Kass95}
Kass, R.~E., \& Raftery, A.~E. 1995,
  \href{http://www.jstor.org/stable/2291091}{\JournalTitle{Journal of the
  American Statistical Association}, 90, 773}

\bibitem[{{Kataria} {et~al.}(2016){Kataria}, {Sing}, {Lewis}, {Visscher},
  {Showman}, {Fortney}, \& {Marley}}]{Kata16}
{Kataria}, T., {Sing}, D.~K., {Lewis}, N.~K., {et~al.} 2016,
  \href{http://dx.doi.org/10.3847/0004-637X/821/1/9}{\JournalTitle{\apj}, 821,
  9}

\bibitem[{{Kellogg} {et~al.}(2016){Kellogg}, {Metchev}, {Gagn{\'e}}, \&
  {Faherty}}]{Kell16}
{Kellogg}, K., {Metchev}, S., {Gagn{\'e}}, J., \& {Faherty}, J. 2016,
  \href{http://dx.doi.org/10.3847/2041-8205/821/1/L15}{\JournalTitle{\apjl},
  821, L15}

\bibitem[{{Kendall} {et~al.}(2004){Kendall}, {Delfosse}, {Mart{\'\i}n}, \&
  {Forveille}}]{Kend04}
{Kendall}, T.~R., {Delfosse}, X., {Mart{\'\i}n}, E.~L., \& {Forveille}, T.
  2004,
  \href{http://dx.doi.org/10.1051/0004-6361:20040046}{\JournalTitle{\aap}, 416,
  L17}

\bibitem[{{Kirkpatrick}(2005)}]{Kirk05}
{Kirkpatrick}, J.~D. 2005,
  \href{http://dx.doi.org/10.1146/annurev.astro.42.053102.134017}{\JournalTitle{\araa},
  43, 195}

\bibitem[{{Kirkpatrick} {et~al.}(1999){Kirkpatrick}, {Allard}, {Bida},
  {Zuckerman}, {Becklin}, {Chabrier}, \& {Baraffe}}]{Kirk99-GD165B}
{Kirkpatrick}, J.~D., {Allard}, F., {Bida}, T., {et~al.} 1999,
  \href{http://dx.doi.org/10.1086/307380}{\JournalTitle{\apj}, 519, 834}

\bibitem[{{Kirkpatrick} {et~al.}(2006){Kirkpatrick}, {Barman}, {Burgasser},
  {McGovern}, {McLean}, {Tinney}, \& {Lowrance}}]{Kirk06}
{Kirkpatrick}, J.~D., {Barman}, T.~S., {Burgasser}, A.~J., {et~al.} 2006,
  \href{http://dx.doi.org/10.1086/499622}{\JournalTitle{\apj}, 639, 1120}

\bibitem[{{Kirkpatrick} {et~al.}(1993){Kirkpatrick}, {Henry}, \&
  {Liebert}}]{Kirk93}
{Kirkpatrick}, J.~D., {Henry}, T.~J., \& {Liebert}, J. 1993,
  \href{http://dx.doi.org/10.1086/172480}{\JournalTitle{\apj}, 406, 701}

\bibitem[{{Kirkpatrick} {et~al.}(2000){Kirkpatrick}, {Reid}, {Liebert},
  {Gizis}, {Burgasser}, {Monet}, {Dahn}, {Nelson}, \& {Williams}}]{Kirk00}
{Kirkpatrick}, J.~D., {Reid}, I.~N., {Liebert}, J., {et~al.} 2000,
  \href{http://dx.doi.org/10.1086/301427}{\JournalTitle{\aj}, 120, 447}

\bibitem[{{Kirkpatrick} {et~al.}(2008){Kirkpatrick}, {Cruz}, {Barman},
  {Burgasser}, {Looper}, {Tinney}, {Gelino}, {Lowrance}, {Liebert},
  {Carpenter}, {Hillenbrand}, \& {Stauffer}}]{Kirk08}
{Kirkpatrick}, J.~D., {Cruz}, K.~L., {Barman}, T.~S., {et~al.} 2008,
  \href{http://dx.doi.org/10.1086/592768}{\JournalTitle{\apj}, 689, 1295}

\bibitem[{{Kirkpatrick} {et~al.}(2016){Kirkpatrick}, {Kellogg}, {Schneider},
  {Fajardo-Acosta}, {Cushing}, {Greco}, {Mace}, {Gelino}, {Wright},
  {Eisenhardt}, {Stern}, {Faherty}, {Sheppard}, {Lansbury}, {Logsdon},
  {Martin}, {McLean}, {Schurr}, {Cutri}, \& {Conrow}}]{Kirk16}
{Kirkpatrick}, J.~D., {Kellogg}, K., {Schneider}, A.~C., {et~al.} 2016,
  \href{http://dx.doi.org/10.3847/0067-0049/224/2/36}{\JournalTitle{\apjs},
  224, 36}

\bibitem[{{Kitzmann} {et~al.}(2020){Kitzmann}, {Heng}, {Oreshenko}, {Grimm},
  {Apai}, {Bowler}, {Burgasser}, \& {Marley}}]{Kitz20}
{Kitzmann}, D., {Heng}, K., {Oreshenko}, M., {et~al.} 2020,
  \href{http://dx.doi.org/10.3847/1538-4357/ab6d71}{\JournalTitle{\apj}, 890,
  174}

\bibitem[{{Koen}(2013)}]{Koen13}
{Koen}, C. 2013,
  \href{http://dx.doi.org/10.1093/mnras/sts208}{\JournalTitle{\mnras}, 428,
  2824}

\bibitem[{{Kumar}(1963)}]{Kumar63a}
{Kumar}, S.~S. 1963,
  \href{http://dx.doi.org/10.1086/147589}{\JournalTitle{\apj}, 137, 1121}

\bibitem[{{Leggett} {et~al.}(2000){Leggett}, {Geballe}, {Fan}, {Schneider},
  {Gunn}, {Lupton}, {Knapp}, {Strauss}, {McDaniel}, {Golimowski}, {Henry},
  {Peng}, {Tsvetanov}, {Uomoto}, {Zheng}, {Hill}, {Ramsey}, {Anderson},
  {Annis}, {Bahcall}, {Brinkmann}, {Chen}, {Csabai}, {Fukugita}, {Hennessy},
  {Hindsley}, {Ivezi{\'c}}, {Lamb}, {Munn}, {Pier}, {Schlegel}, {Smith},
  {Stoughton}, {Thakar}, \& {York}}]{Legg00}
{Leggett}, S.~K., {Geballe}, T.~R., {Fan}, X., {et~al.} 2000,
  \href{http://dx.doi.org/10.1086/312728}{\JournalTitle{\apjl}, 536, L35}

\bibitem[{{Leggett} {et~al.}(2002){Leggett}, {Golimowski}, {Fan}, {Geballe},
  {Knapp}, {Brinkmann}, {Csabai}, {Gunn}, {Hawley}, {Henry}, {Hindsley},
  {Ivezi{\'c}}, {Lupton}, {Pier}, {Schneider}, {Smith}, {Strauss}, {Uomoto}, \&
  {York}}]{Legg02a}
{Leggett}, S.~K., {Golimowski}, D.~A., {Fan}, X., {et~al.} 2002,
  \href{http://dx.doi.org/10.1086/324037}{\JournalTitle{\apj}, 564, 452}

\bibitem[{{Lindegren, L.} {et~al.}(2018){Lindegren, L.}, {Hernandez, J.},
  {Bombrun, A.}, {Klioner, S.}, {Bastian, U.}, \& {Ramos-Lerate, M.}}]{Lind18}
{Lindegren, L.}, {Hernandez, J.}, {Bombrun, A.}, {et~al.} 2018,
  \href{http://dx.doi.org/10.1051/0004-6361/201832727}{\JournalTitle{A\&A}}

\bibitem[{{Line} {et~al.}(2014){Line}, {Knutson}, {Wolf}, \& {Yung}}]{Line14}
{Line}, M.~R., {Knutson}, H., {Wolf}, A.~S., \& {Yung}, Y.~L. 2014,
  \href{http://dx.doi.org/10.1088/0004-637X/783/2/70}{\JournalTitle{\apj}, 783,
  70}

\bibitem[{{Line} {et~al.}(2015){Line}, {Teske}, {Burningham}, {Fortney}, \&
  {Marley}}]{Line15}
{Line}, M.~R., {Teske}, J., {Burningham}, B., {Fortney}, J.~J., \& {Marley},
  M.~S. 2015,
  \href{http://dx.doi.org/10.1088/0004-637X/807/2/183}{\JournalTitle{\apj},
  807, 183}

\bibitem[{{Line} {et~al.}(2017){Line}, {Marley}, {Liu}, {Burningham}, {Morley},
  {Hinkel}, {Teske}, {Fortney}, {Freedman}, \& {Lupu}}]{Line17}
{Line}, M.~R., {Marley}, M.~S., {Liu}, M.~C., {et~al.} 2017,
  \href{http://dx.doi.org/10.3847/1538-4357/aa7ff0}{\JournalTitle{\apj}, 848,
  83}

\bibitem[{{Lodders}(1999)}]{Lodd99}
{Lodders}, K. 1999,
  \href{http://dx.doi.org/10.1086/307387}{\JournalTitle{\apj}, 519, 793}

\bibitem[{{Lodders}(2002)}]{Lodd02}
---. 2002, \href{http://dx.doi.org/10.1086/342241}{\JournalTitle{\apj}, 577,
  974}

\bibitem[{{Lodders}(2010)}]{Lodd10}
---. 2010, {Exoplanet Chemistry} ({Wiley}), 157

\bibitem[{{Lodders} \& {Fegley}(2002)}]{Lodd02b}
{Lodders}, K., \& {Fegley}, B. 2002,
  \href{http://dx.doi.org/10.1006/icar.2001.6740}{\JournalTitle{Icarus}, 155,
  393}

\bibitem[{{Lodders} \& {Fegley}(2006)}]{Lodd06}
{Lodders}, K., \& {Fegley}, Jr., B. 2006, {Chemistry of Low Mass Substellar
  Objects} ({Praxis Publishing Ltd, Chichester, UK}), 1

\bibitem[{{Lueber} {et~al.}(2022){Lueber}, {Kitzmann}, {Bowler}, {Burgasser},
  \& {Heng}}]{Lueb22}
{Lueber}, A., {Kitzmann}, D., {Bowler}, B.~P., {Burgasser}, A.~J., \& {Heng},
  K. 2022, \JournalTitle{arXiv e-prints}, arXiv:2204.01330

\bibitem[{{Madhusudhan} \& {Seager}(2009)}]{Madh09}
{Madhusudhan}, N., \& {Seager}, S. 2009,
  \href{http://dx.doi.org/10.1088/0004-637X/707/1/24}{\JournalTitle{\apj}, 707,
  24}

\bibitem[{{Marley} {et~al.}(2021){Marley}, {Saumon}, {Visscher}, {Lupu},
  {Freedman}, {Morley}, {Fortney}, {Seay}, {Smith}, {Teal}, \& {Wang}}]{Sonora}
{Marley}, M.~S., {Saumon}, D., {Visscher}, C., {et~al.} 2021,
  \href{http://dx.doi.org/10.3847/1538-4357/ac141d}{\JournalTitle{\apj}, 920,
  85}

\bibitem[{{Martin} {et~al.}(2017){Martin}, {Mace}, {McLean}, {Logsdon}, {Rice},
  {Kirkpatrick}, {Burgasser}, {McGovern}, \& {Prato}}]{Martin17}
{Martin}, E.~C., {Mace}, G.~N., {McLean}, I.~S., {et~al.} 2017,
  \href{http://dx.doi.org/10.3847/1538-4357/aa6338}{\JournalTitle{\apj}, 838,
  73}

\bibitem[{{McBride} \& {Gordon}(1994)}]{McBr94}
{McBride}, B., \& {Gordon}, S. 1994, {Computer Program for Calculation of
  Complex Chemical Equilibrium Compositions and Applications I. Analysis},
  Reference Publication NASA RP-1311, NASA Lewis Research Center, National
  Aeronautics and Space Administration Lewis Research Center Cleveland, Ohio
  44135-319, https://www.grc.nasa.gov/WWW/CEAWeb/RP-1311.htm

\bibitem[{{McLean} {et~al.}(2007){McLean}, {Prato}, {McGovern}, {Burgasser},
  {Kirkpatrick}, {Rice}, \& {Kim}}]{McLe07}
{McLean}, I.~S., {Prato}, L., {McGovern}, M.~R., {et~al.} 2007,
  \href{http://dx.doi.org/10.1086/511740}{\JournalTitle{\apj}, 658, 1217}

\bibitem[{Miles-P{\'{a}}ez {et~al.}(2017)Miles-P{\'{a}}ez, Metchev, Heinze, \&
  Apai}]{Miles17}
Miles-P{\'{a}}ez, P.~A., Metchev, S.~A., Heinze, A., \& Apai, D. 2017,
  \href{http://dx.doi.org/10.3847/1538-4357/aa6f11}{\JournalTitle{The
  Astrophysical Journal}, 840, 83}

\bibitem[{{Mohanty} \& {Basri}(2003)}]{Moha03}
{Mohanty}, S., \& {Basri}, G. 2003,
  \href{http://dx.doi.org/10.1086/345097}{\JournalTitle{\apj}, 583, 451}

\bibitem[{{Morley} {et~al.}(2013){Morley}, {Fortney}, {Kempton}, {Marley},
  {Visscher}, \& {Zahnle}}]{Morl13}
{Morley}, C.~V., {Fortney}, J.~J., {Kempton}, E.~M.-R., {et~al.} 2013,
  \href{http://dx.doi.org/10.1088/0004-637X/775/1/33}{\JournalTitle{\apj}, 775,
  33}

\bibitem[{{Morley} {et~al.}(2012){Morley}, {Fortney}, {Marley}, {Visscher},
  {Saumon}, \& {Leggett}}]{Morl12}
{Morley}, C.~V., {Fortney}, J.~J., {Marley}, M.~S., {et~al.} 2012,
  \href{http://dx.doi.org/10.1088/0004-637X/756/2/172}{\JournalTitle{\apj},
  756, 172}

\bibitem[{{Moses} {et~al.}(2012){Moses}, {Richardson}, {Madhusudhan}, {Line},
  {Visscher}, \& {Fortney}}]{Moses12}
{Moses}, J.~I., {Richardson}, M.~R., {Madhusudhan}, N., {et~al.} 2012, in
  AAS/Division for Planetary Sciences Meeting Abstracts, Vol.~44, AAS/Division
  for Planetary Sciences Meeting Abstracts, 103.02

\bibitem[{{Moses} {et~al.}(2013){Moses}, {Line}, {Visscher}, {Richardson},
  {Nettelmann}, {Fortney}, {Barman}, {Stevenson}, \& {Madhusudhan}}]{Moses13}
{Moses}, J.~I., {Line}, M.~R., {Visscher}, C., {et~al.} 2013,
  \href{http://dx.doi.org/10.1088/0004-637X/777/1/34}{\JournalTitle{\apj}, 777,
  34}

\bibitem[{{Patten} {et~al.}(2006){Patten}, {Stauffer}, {Burrows}, {Marengo},
  {Hora}, {Luhman}, {Sonnett}, {Henry}, {Raghavan}, {Megeath}, {Liebert}, \&
  {Fazio}}]{Patt06}
{Patten}, B.~M., {Stauffer}, J.~R., {Burrows}, A., {et~al.} 2006,
  \href{http://dx.doi.org/10.1086/507264}{\JournalTitle{\apj}, 651, 502}

\bibitem[{{Piette} \& {Madhusudhan}(2020)}]{Piet20}
{Piette}, A. A.~A., \& {Madhusudhan}, N. 2020,
  \href{http://dx.doi.org/10.1093/mnras/staa2289}{\JournalTitle{\mnras}, 497,
  5136}

\bibitem[{{Prato} {et~al.}(2015){Prato}, {Mace}, {Rice}, {McLean},
  {Kirkpatrick}, {Burgasser}, \& {Kim}}]{Prat15}
{Prato}, L., {Mace}, G.~N., {Rice}, E.~L., {et~al.} 2015,
  \href{http://dx.doi.org/10.1088/0004-637X/808/1/12}{\JournalTitle{\apj}, 808,
  12}

\bibitem[{{Reid} {et~al.}(2008){Reid}, {Cruz}, {Kirkpatrick}, {Allen},
  {Mungall}, {Liebert}, {Lowrance}, \& {Sweet}}]{Reid08b}
{Reid}, I.~N., {Cruz}, K.~L., {Kirkpatrick}, J.~D., {et~al.} 2008,
  \href{http://dx.doi.org/10.1088/0004-6256/136/3/1290}{\JournalTitle{\aj},
  136, 1290}

\bibitem[{{Saumon} {et~al.}(1996){Saumon}, {Hubbard}, {Burrows}, {Guillot},
  {Lunine}, \& {Chabrier}}]{Saum96}
{Saumon}, D., {Hubbard}, W.~B., {Burrows}, A., {et~al.} 1996,
  \href{http://dx.doi.org/10.1086/177027}{\JournalTitle{\apj}, 460, 993}

\bibitem[{{Saumon} \& {Marley}(2008)}]{Saum08}
{Saumon}, D., \& {Marley}, M.~S. 2008,
  \href{http://dx.doi.org/10.1086/592734}{\JournalTitle{\apj}, 689, 1327}

\bibitem[{{Schmidt} {et~al.}(2010){Schmidt}, {West}, {Burgasser}, {Bochanski},
  \& {Hawley}}]{Schm10a}
{Schmidt}, S.~J., {West}, A.~A., {Burgasser}, A.~J., {Bochanski}, J.~J., \&
  {Hawley}, S.~L. 2010,
  \href{http://dx.doi.org/10.1088/0004-6256/139/3/1045}{\JournalTitle{\aj},
  139, 1045}

\bibitem[{{Schneider} {et~al.}(2014){Schneider}, {Cushing}, {Kirkpatrick},
  {Mace}, {Gelino}, {Faherty}, {Fajardo-Acosta}, \& {Sheppard}}]{Schn14}
{Schneider}, A.~C., {Cushing}, M.~C., {Kirkpatrick}, J.~D., {et~al.} 2014,
  \href{http://dx.doi.org/10.1088/0004-6256/147/2/34}{\JournalTitle{\aj}, 147,
  34}

\bibitem[{Sheppard \& Cushing(2008)}]{Shep08}
Sheppard, S.~S., \& Cushing, M.~C. 2008,
  \href{http://dx.doi.org/10.1088/0004-6256/137/1/304}{\JournalTitle{The
  Astronomical Journal}, 137, 304}

\bibitem[{{Skemer} {et~al.}(2016){Skemer}, {Morley}, {Zimmerman}, {Skrutskie},
  {Leisenring}, {Buenzli}, {Bonnefoy}, {Bailey}, {Hinz}, {Defr{\'e}re},
  {Esposito}, {Apai}, {Biller}, {Brandner}, {Close}, {Crepp}, {De Rosa},
  {Desidera}, {Eisner}, {Fortney}, {Freedman}, {Henning}, {Hofmann},
  {Kopytova}, {Lupu}, {Maire}, {Males}, {Marley}, {Morzinski}, {Oza},
  {Patience}, {Rajan}, {Rieke}, {Schertl}, {Schlieder}, {Stone}, {Su}, {Vaz},
  {Visscher}, {Ward-Duong}, {Weigelt}, \& {Woodward}}]{Skem16}
{Skemer}, A.~J., {Morley}, C.~V., {Zimmerman}, N.~T., {et~al.} 2016,
  \href{http://dx.doi.org/10.3847/0004-637X/817/2/166}{\JournalTitle{\apj},
  817, 166}

\bibitem[{{Sorahana} {et~al.}(2013){Sorahana}, {Yamamura}, \&
  {Murakami}}]{Sora13}
{Sorahana}, S., {Yamamura}, I., \& {Murakami}, H. 2013,
  \href{http://dx.doi.org/10.1088/0004-637X/767/1/77}{\JournalTitle{\apj}, 767,
  77}

\bibitem[{{Tinney} {et~al.}(1995){Tinney}, {Reid}, {Gizis}, \&
  {Mould}}]{Tinn95}
{Tinney}, C.~G., {Reid}, I.~N., {Gizis}, J., \& {Mould}, J.~R. 1995,
  \href{http://dx.doi.org/10.1086/117743}{\JournalTitle{\aj}, 110, 3014}

\bibitem[{{van Altena} {et~al.}(1995){van Altena}, {Lee}, \&
  {Hoffleit}}]{vanA95}
{van Altena}, W.~F., {Lee}, J.~T., \& {Hoffleit}, E.~D. 1995, {The general
  catalogue of trigonometric [stellar] parallaxes}

\bibitem[{{Visscher}(2012)}]{Viss12}
{Visscher}, C. 2012,
  \href{http://dx.doi.org/10.1088/0004-637X/757/1/5}{\JournalTitle{\apj}, 757,
  5}

\bibitem[{{Visscher} {et~al.}(2006){Visscher}, {Lodders}, \& {Fegley}}]{Viss06}
{Visscher}, C., {Lodders}, K., \& {Fegley}, Jr., B. 2006,
  \href{http://dx.doi.org/10.1086/506245}{\JournalTitle{\apj}, 648, 1181}

\bibitem[{{Visscher} {et~al.}(2010){Visscher}, {Lodders}, \&
  {Fegley}}]{Viss10a}
---. 2010,
  \href{http://dx.doi.org/10.1088/0004-637X/716/2/1060}{\JournalTitle{\apj},
  716, 1060}

\bibitem[{{Vrba} {et~al.}(2004){Vrba}, {Henden}, {Luginbuhl}, {Guetter},
  {Munn}, {Canzian}, {Burgasser}, {Kirkpatrick}, {Fan}, {Geballe},
  {Golimowski}, {Knapp}, {Leggett}, {Schneider}, \& {Brinkmann}}]{Vrba04}
{Vrba}, F.~J., {Henden}, A.~A., {Luginbuhl}, C.~B., {et~al.} 2004,
  \href{http://dx.doi.org/10.1086/383554}{\JournalTitle{\aj}, 127, 2948}

\bibitem[{{Wakeford} {et~al.}(2017){Wakeford}, {Visscher}, {Lewis}, {Kataria},
  {Marley}, {Fortney}, \& {Mand ell}}]{Wake17}
{Wakeford}, H.~R., {Visscher}, C., {Lewis}, N.~K., {et~al.} 2017,
  \href{http://dx.doi.org/10.1093/mnras/stw2639}{\JournalTitle{\mnras}, 464,
  4247}

\bibitem[{Yamamura {et~al.}(2010)Yamamura, Tsuji, \& Tanab{\'{e}}}]{Yama10}
Yamamura, I., Tsuji, T., \& Tanab{\'{e}}, T. 2010,
  \href{http://dx.doi.org/10.1088/0004-637x/722/1/682}{\JournalTitle{The
  Astrophysical Journal}, 722, 682}

\bibitem[{{Zalesky} {et~al.}(2019){Zalesky}, {Line}, {Schneider}, \&
  {Patience}}]{Zale19}
{Zalesky}, J.~A., {Line}, M.~R., {Schneider}, A.~C., \& {Patience}, J. 2019,
  \href{http://dx.doi.org/10.3847/1538-4357/ab16db}{\JournalTitle{\apj}, 877,
  24}

\bibitem[{{Zapatero Osorio} {et~al.}(2005){Zapatero Osorio}, {Caballero}, \&
  {B{\'e}jar}}]{Zapa05}
{Zapatero Osorio}, M.~R., {Caballero}, J.~A., \& {B{\'e}jar}, V.~J.~S. 2005,
  \href{http://dx.doi.org/10.1086/427433}{\JournalTitle{\apj}, 621, 445}

\bibitem[{{Zhang} {et~al.}(2017){Zhang}, {Pinfield}, {G{\'a}lvez-Ortiz},
  {Burningham}, {Lodieu}, {Marocco}, {Burgasser}, {Day-Jones}, {Allard},
  {Jones}, {Homeier}, {Gomes}, \& {Smart}}]{Zhang2017a}
{Zhang}, Z.~H., {Pinfield}, D.~J., {G{\'a}lvez-Ortiz}, M.~C., {et~al.} 2017,
  \href{http://dx.doi.org/10.1093/mnras/stw2438}{\JournalTitle{\mnras}, 464,
  3040}

\end{thebibliography}

\end{document}